\documentclass[10pt,aps,prd,onecolumn,notitlepage,nofootinbib,floatfix,superscriptaddress]{revtex4-1}

\usepackage{amsmath, amsthm, amssymb, amsfonts, amsbsy, mathrsfs}

\usepackage{url}
\usepackage{xcolor}
\usepackage{calligra,bm}
\usepackage{stmaryrd}
\usepackage{dcolumn}
\usepackage{varioref}
\usepackage{textcomp}

\allowdisplaybreaks
\usepackage[colorlinks,citecolor=blue,urlcolor=magenta,linkcolor=blue,bookmarks=true]{hyperref} %make sure it comes last of your loaded packages
\hypersetup{pdfstartview=FitH,pdfhighlight=/O}

\def\bg{\bar{g}}
\def\bd{\bar{\nabla}}

\labelformat{section}{Sec. #1}
\labelformat{subsection}{Sec. #1}
\labelformat{subsubsection}{Sec. #1}
\labelformat{subsubsubsection}{Sec. #1}
\labelformat{equation}{(#1)}
\labelformat{figure}{Figure~#1}
\labelformat{subfigure}{Fig.~\thefigure#1}
\labelformat{table}{Tab.~#1}
\labelformat{appendix}{Appendix #1}
\numberwithin{equation}{section}

\begin{document}

\title{Perturbations of the almost Killing equation and their implications}

\author{Sumanta Chakraborty}
%  \affiliation{School of Mathematical and Computational Sciences, Indian Association for the Cultivation of Science, Kolkata-700032, India}
\affiliation{School of Physical Sciences, Indian Association for the Cultivation of Science, Kolkata-700032, India}

\author{Justin C. Feng}
\affiliation{CENTRA, Departamento de F{\'i}sica, Instituto Superior T{\'e}cnico – IST, Universidade de Lisboa – UL, Avenida Rovisco Pais 1, 1049 Lisboa, Portugal}

%\preprint{}

%\date{\today}

%-----------------------------------------------------------------------
%-----------------------------------
%-----------------
%--------
%---
%-
%
%
%-
%---
%--------
%-----------------
%-----------------------------------
%-----------------------------------------------------------------------

%=======================================================================
%-----------------------------------------------------------------------
%
%		ABSTRACT
%
%-----------------------------------------------------------------------
%=======================================================================
\begin{abstract}
Killing vectors play a crucial role in characterizing the symmetries of a given spacetime. However, realistic astrophysical systems are in most cases only approximately symmetric. Even in the case of an astrophysical black hole, one might expect Killing symmetries to exist only in an approximate sense due to perturbations from external matter fields. In this work, we consider the generalized notion of Killing vectors provided by the almost Killing equation, and study the perturbations induced by a perturbation of a background spacetime satisfying exact Killing symmetry. To first order, we demonstrate that for nonradiative metric perturbations (that is, metric perturbations with nonvanishing trace) of symmetric vacuum spacetimes, the perturbed almost Killing equation avoids the problem of an unbounded Hamiltonian for hyperbolic parameter choices. For traceless metric perturbations, we obtain similar results for the second-order perturbation of the almost Killing equation, with some additional caveats. Thermodynamical implications are also explored.
\end{abstract}

% insert suggested PACS numbers in braces on next line
%\pacs{}
% insert suggested keywords - APS authors don't need to do this
%\keywords{}

%-----------------------------------------------------------------------
%-----------------------------------
%-----------------
%--------
%---
%-
%
%
%-
%---
%--------
%-----------------
%-----------------------------------
%-----------------------------------------------------------------------

\maketitle

%-----------------------------------------------------------------------
%-----------------------------------
%-----------------
%--------
%---
%-
%
%
%-
%---
%--------
%-----------------
%-----------------------------------
%-----------------------------------------------------------------------

%=======================================================================
%-----------------------------------------------------------------------
%
%		INTRODUCTION
%
%-----------------------------------------------------------------------
%=======================================================================

\section{Introduction}

Symmetries are central to our understanding of the physical world and play a key role in describing a wide range of physical systems, from the determination of the Lagrangian for a mechanical system to the lattice structure of crystalline substances. This extends to general relativity and relativistic theories of gravity: Symmetries and the Killing vectors that formalize them are useful for characterizing the properties of spacetime and matter. For example, the existence of a timelike Killing vector field ensures that the spacetime is time-translation invariant, leading to conserved definitions for energy for test particles and matter. Similarly, the existence of a closed spacelike Killing vector field ensures rotational invariance of the spacetime geometry, leading to a conserved definition for angular momentum. Moreover, many astrophysical systems are approximately described by spacetime geometries admitting such Killing vectors. However, the exact nature of these symmetries is lost in realistic systems due to dynamical behavior of and irregularities in the matter configurations. This scenario can arise in various contexts; e.g., when one drops a cup of coffee into a black hole (considering the gravitational backreaction), the resulting perturbed spacetime no longer inherits the exact Killing symmetry. Fortunately, one may still construct certain generalizations of Killing vector fields in such circumstances, which are useful for understanding generalizations of conserved quantities (such as energy and momentum) for gravitating systems that lack exact symmetries.

The literature contains several approaches for defining generalized Killing vectors and symmetries. Specific examples include Matzner's eigenvector approach \cite{Matzner1968,*Beetle2008,*BeetleWilder2014}, which has recently been of interest for studying quantum geometries in causal dynamical triangulations \cite{BrunekreefReitz2020}, symmetry-seeking coordinates \cite{GarfinkleGundlach1999}, affine collineations \cite{Harte2008}, and the almost Killing equation (henceforth, AKE) \cite{Taubes1978,Bonaetal2005}. The latter approach, the generalized Killing vectors defined by the AKE, forms the subject of this paper.

The generalized Killing vector fields (henceforth GKVs) associated with the AKE may be used to define conserved charges in spacetimes with no exact Killing symmetries. Given some notion of the GKV, the generalized Komar current, as defined in \cite{Komar1959,Komar1962}, may be used to construct generalizations of the usual Komar charges---explicit examples have been constructed and studied in \cite{Ruizetal2014,FengCurrents2018} (see also \cite{Peng2020,*Peng2019} for further generalizations of the Komar current). For example, in \cite{FGW2019}, it was shown that the generalized Komar current for solutions of the AKE, which are the GKVs, can provide a measure of the matter content of the physical system under consideration. It was also demonstrated in \cite{FGW2019} that GKVs may be used with the generalized Komar current to obtain a Gauss law for systems of black holes in vacuum and matter distributions with compact support if the GKV is divergenceless or for a certain choice of parameters associated with the AKE.

Though one might hope that, for sufficiently small perturbations of symmetric spacetimes, the solutions of the AKE are close to that of an exact Killing vector field, GKVs do not necessarily approximate Killing vectors in the sense that the components of $\nabla_{(\mu} \chi_{\nu)}$ can be large compared to that of $\chi^\alpha$ (where $\chi^{\alpha}$ is a GKV), even in Minkowski spacetime. One might postulate that an appropriate choice for initial data for the AKE will ensure that $\nabla_{(\mu} \chi_{\nu)}$ is small in the sense that the components of $\nabla_{(\mu} \chi_{\nu)}$ are much smaller than that of $\chi^\alpha$ for some normalization. This matter was studied to some degree in \cite{FGW2019}, which examines the hyperbolicity and Hamiltonian stability of the system described by the AKE. There, parameter choices were identified in which the AKE is strongly and weakly hyperbolic, and also in which the system admits ghost modes and unbounded Hamiltonians. Ghosts and unbounded Hamiltonians are potentially dangerous, as they may correspond to runaway behavior which can potentially drive the solution far from the Killing condition
$\nabla_{(\mu} \chi_{\nu)} = 0$, even if the initial data approximately satisfy this condition (and its time derivative). Though there is no parameter choice for which the generic AKE system is both hyperbolic and has a bounded Hamiltonian, it was shown in \cite{FGW2019} that in vacuum ($R_{\mu \nu} = 0$) spacetime and for initial data satisfying $\nabla \cdot \chi = 0$ and its derivative, the system yields a constraint which renders it dynamically equivalent to a system with a bounded Hamiltonian and simultaneously equivalent to a strongly hyperbolic system. Moreover, it was argued that for appropriate initial data and falloff conditions, the AKE can provide a notion of an approximate Killing vector in a neighborhood of spatial infinity of asymptotically flat spacetimes.

Despite the promising results presented in \cite{FGW2019} for the vacuum case, these do not in general extend to the nonvacuum ($R_{\mu \nu} \neq 0$) case. Therefore, it is not immediately apparent that the AKE can be simultaneously well posed and equivalent to a system with a bounded Hamiltonian for spacetimes containing matter. On the other hand, for perturbations of spacetimes that admit an exact Killing vector, one might expect the AKE for the perturbed spacetime to admit solutions that approximate Killing vectors. Thus, one of the primary aims of this article is to construct perturbative solutions to the AKE for perturbations of spacetimes which admit Killing vectors and to study their properties and the interpretation of the resulting generalized Komar currents and charges. Additionally, we would like to explore the connection of the perturbed Komar current and charges with the thermodynamic behavior of black hole spacetimes, e.g., the first law. As we will show, there is a close correspondence between the AKE and black hole thermodynamics.

The paper is organized as follows: In Eq.~\ref{AKE_review} we will review the AKE and shall present a physically interesting scenario, namely, that of the Vaidya spacetime, where some of the key aspects of the AKE will be demonstrated. Subsequently, the evolution of the GKVs in the perturbed spacetime will be presented in Eq.~\ref{SGKVP} from both the action formalism and also from the perturbation of the AKE itself. The stability of the perturbed AKE, as well as its hyperbolicity, will be studied in Eq.~\ref{pake_hamiltonian}, before discussing the nature of the solution of the AKE for both first- and second-order perturbations in Eq.~\ref{pake_sol}. Finally, the thermodynamic interpretation of the AKE will be depicted in Eq.~\ref{pake_thermodynamics}, before presenting the concluding remarks in Eq.~\ref{pake_conclusions}.
%-unc Several computations performed in this work have been presented in detail in \ref{AppA} --- \ref{AppF}.

\emph{Notations and conventions:} Throughout this paper, we will use the mostly positive signature convention, such that the Minkowski metric in the Cartesian coordinates has the following form: $\eta_{\mu \nu}=\textrm{diag.}(-1,1,1,1)$. The four-dimensional spacetime indices will be denoted by the greek letters $\mu,\nu,\alpha,\ldots$. We will work in units such that the fundamental constants have the values $G=c=\hbar=1$. Throughout the article, indices on quantities which appear in arguments will be denoted with superscript and subscript dots; for instance, the arguments in $\mathcal{A}[\chi^\cdot]$ and $\mathcal{O}(h_{\cdot \cdot}$ represent the quantities $\chi^\mu$ and $h_{\mu \nu}$.

%=======================================================================
%-----------------------------------------------------------------------
%
%		ALMOST KILLING EQUATION
%
%-----------------------------------------------------------------------
%=======================================================================
\section{The Almost Killing Equation: A brief review}\label{AKE_review}

In this section, we will briefly review the almost Killing equation, where the motivation for its construction and its various properties will be discussed in detail. In addition, we will also present the Vaidya geometry as an example of arriving at a solution of the almost Killing equation.

%==============================================
%-----------------------------------------------------------------------
%		MOTIVATION AND PROPERTY
%-----------------------------------------------------------------------
%==============================================
\subsection{Motivation, construction, and properties}

A Killing vector field $\xi^{\mu}$ is defined as one which satisfies the Killing equation $\pounds_\xi g_{\mu \nu} = 2\nabla_{(\mu} \xi_{\nu)} = 0$. The divergence of the Killing equation takes the form
%===============================================
\begin{equation} \label{DivKillingEquation}
\Box \xi^{\alpha} + R{^\alpha}{_\beta} \, \xi^{\beta} = 0 ~.
\end{equation}
%===============================================
As evident, Eq.~\ref{DivKillingEquation} takes the form of a wave equation; on geometries that do not admit Killing vectors, one can nonetheless construct generalizations of the Killing equation by solving Eq.~\ref{DivKillingEquation} for an appropriate set of initial data. The AKE is a generalization of Eq.~\ref{DivKillingEquation}, and is given by the following formula:
%===============================================
\begin{equation} \label{AlmostKillingEquation}
\Box \chi^{\alpha} + R{^\alpha}{_\beta} \, \chi^{\beta} +  \nabla^\alpha \left[(1 - \mu)\nabla \cdot \chi \right]= 0 ~,
\end{equation}
%===============================================
where $\mu$ is a scalar, which in previous literature is assumed to be a constant; for generality, we do not assume this to be the case here. The solution of Eq.~\ref{AlmostKillingEquation}, i.e., $\chi^{\alpha}$, is the GKV. It is straightforward to verify that Killing vectors satisfy the AKE; it is in this sense that solutions of the AKE may be regarded as generalizations of Killing vectors. As discussed in \cite{FGW2019}, GKVs are not necessarily approximate Killing vectors in the following sense. The vector $\chi^\alpha$ satisfies the AKE if the tensor $Q_{\mu \nu} = \nabla_{(\mu} \chi_{\nu)}$ is transverse and trace-free; however, the components of transverse and trace-free tensors need not be small.\footnote{Alternatively, one can show that even on a Minkowskian background, the AKE admits wavelike solutions for which the components $Q_{\mu \nu} \gg 0$, so that they cannot be considered as approximate Killing vectors by any means.}

It is instructive to derive any evolution equation from an action principle, and the AKE is no different. The AKE presented in Eq.~\ref{AlmostKillingEquation} may be derived from the following action (see the Appendix for a derivation of the AKE from this action functional):
%===============================================
\begin{equation}\label{ActionComp}
\mathcal{A}[\chi^\cdot]= \int_{\mathcal{M}}d^{4}x\sqrt{-g} \left(-{\nabla}^{(\alpha} {\chi}^{\beta)} \, {\nabla}_{(\alpha} {\chi}_{\beta)} + \frac{1}{2} {\mu} ({\nabla} \cdot {\chi})^2 \right)~.
\end{equation}
%===============================================
Here, $\mathcal{M}$ denotes the spacetime volume of interest, and as evident, it immediately follows that the action $\mathcal{A}[\chi^\cdot]$ vanishes if $\chi^{\alpha}$ is a Killing vector field. It was shown in \cite{FGW2019} that the AKE is strongly hyperbolic when the parameter $\mu=1$; however, it fails to be hyperbolic when $\mu \neq 2$, and is weakly hyperbolic for all other (constant) values for $\mu$. It was also argued that in general, the AKE may suffer from dynamical instabilities; a Hamiltonian analysis reveals the presence of ghosts for $\mu<2$, and unbounded terms for $\mu>1/3$. There is no parameter choice for $\mu$ in which the AKE avoids these potential instabilities and is also hyperbolic. However, for vacuum spacetimes, at least one exception exists, which we will discuss shortly.

In any spacetime manifold, given some vector field $V^\alpha$, it follows from differential geometry that it is possible to construct a conserved current and hence a conserved charge. This conserved current takes the following form:
%===============================================
\begin{equation} \label{NoetherCurrent}
J^\alpha = \nabla_\beta \left( \nabla^\alpha V^\beta - \nabla^\beta V^\alpha \right)~.
\end{equation}
%===============================================
For Killing vector fields, the conserved current $J^{\alpha}$ is known as the Komar current, and the associated charges are known as the Komar charges, and we have so far referred to these, respectively, as the generalized Komar current and the generalized Komar charges. However, if the vector field $V^\alpha$ is considered as a generator of the diffeomorphism, then $J^\alpha$ is in fact the conserved current corresponding to the invariance under said diffeomorphism.\footnote{For further discussion of this point, see \cite{Katz1985,*Baketal1993,*KBL1997,*Obukhovetal2006a,*Obukhovetal2006b,Deruelleetal2004,*Obukhovetal2006c,*Obukhovetal2013,*Obukhovetal2014,*Obukhovetal2015,*SchmidtBicak2018}.} For this reason, it is perhaps more appropriate to call this the Noether current since it arises out of the diffeomorphism invariance of the gravitational system \cite{Iyer:1994ys,Jacobson:2015uqa,Prabhu:2015vua,Chakraborty:2018qew,Aneesh:2020fcr}. For the remainder of this article, we shall use this terminology and shall refer to the conserved charges associated with $J^\alpha$ as Noether charges. As \cite{Padmanabhan:2013nxa,Chakraborty:2015hna,Chakraborty:2015wma} demonstrate, Noether charges defined in this manner have interesting thermodynamical interpretations when computed over certain spacelike and null surfaces.

As emphasized before, the Noether current $J^{\alpha}$ defined in Eq.~\ref{NoetherCurrent} is identically divergence-free, which when evaluated for solutions of the AKE takes the following form:
%===============================================
\begin{equation} \label{KomarAK}
J^\alpha_{\chi} = 2R{^\alpha}{_\beta} \chi^\beta + \nabla^\alpha \left[(2-\mu) \nabla \cdot \chi \right]~.
\end{equation}
%===============================================
We note that when $\mu=2$, the Noether current may be interpreted as a measure of the energy and momentum through the use of the trace-reversed Einstein field equations:
%===============================================
\begin{equation} \label{EFE}
R_{\alpha \beta} = 8 \pi \left( T_{\alpha \beta} - \frac{1}{2} g_{\alpha \beta} \, T \right)~.
\end{equation}
%===============================================
Moreover, the divergence-free property of the Noether current yields the following expression:
%===============================================
\begin{equation} \label{KomarAKcond}
\Box \left[(\mu - 2)\nabla \cdot \chi \right] = \chi^\beta \nabla_\beta R + 2 R^{\alpha \beta} \nabla_{(\alpha} \chi_{\beta)}~,
\end{equation}
%===============================================
where the contracted Bianchi identity $\nabla_{\alpha}R{^\alpha}_{\beta}=(1/2)\nabla_{\beta}R$ has been used. The above evolution equation for $(\nabla \cdot \chi)$ was used in \cite{FGW2019} to show that in a vacuum spacetime, the constraint $\nabla \cdot \chi = 0$ is propagated by the AKE; if the initial data satisfy the constraint $\nabla \cdot \chi = 0$ and its time derivative, then the time development of the solution satisfies the constraint. Under this constraint, the AKE becomes strongly hyperbolic and is independent of $\mu$, so that it is no longer subject to the instabilities associated with ghosts or unbounded terms in the Hamiltonian.

%=======================================================================
%-----------------------------------------------------------------------
%		EXAMPLE: THE VAIDYA GEOMETRY
%-----------------------------------------------------------------------
%=======================================================================
\subsection{Example: The Vaidya geometry}\label{vaidya_examp}

Here, we review and generalize the solution for the AKE in the Vaidya spacetime, as presented in \cite{FengCurrents2018}, to gain some insight into the relationship between GKVs, the Smarr relation, and the laws of thermodynamics. The line element associated with Vaidya spacetime takes the following form ($d\Omega^2$ being the round metric on the 2-sphere):
%===============================================
\begin{equation}\label{VaidyaLineElem}
ds^2=-\left[1-\frac{2 M(v)}{r}\right] dv^{2} + 2 \, dv \, dr + r^2 d\Omega^2~,
\end{equation}
%===============================================
where $M(v)$ is an arbitrary positive function of the advanced null coordinate $v$. Following \cite{FengCurrents2018}, here also we construct a solution to the AKE in the Vaidya spacetime presented in Eq.~\ref{VaidyaLineElem} for the $\mu=2$ case, which has the following form:
%===============================================
\begin{equation} \label{AKEVaidyaPulse}
\chi^{\alpha}=\left(\frac{M(v)}{M_0},\frac{r \, M^\prime(v) + f(v)}{M_0},0,0\right)~,
\end{equation}
%===============================================
where $f(v)$ is an arbitrary function of the advanced null coordinate $v$, and $1/M_{0}$ is a constant factor; a natural choice for this factor is to set it equal to the ADM mass of the spacetime. The fact that the GKV $\chi^{\alpha}$ depends on an arbitrary function $f(v)$ is due to the fact that for $\mu=2$, the AKE fails to be hyperbolic. Fortunately, the resulting Noether current and Noether charge are unaffected by the arbitrary function $f(v)$, so that one can regard it as a sort of ``gauge'' potential. There is, however, a criterion that one may use to fix this arbitrary function $f(v)$, which will be discussed later. The Noether current associated with the GKV $\chi^{\alpha}$ then takes the following form:
%===============================================
\begin{equation} \label{AKEVaidyaPulseKomar}
J_{\chi}^{\alpha}=\left(0,\frac{4 \, M(v) \, M'(v)}{M_0 \, r^2},0,0\right)~.
\end{equation}
%===============================================
In the Vaidya spacetime, the surface characterized by
%===============================================
\begin{equation}\label{HorizonCond}
r_{\rm H}:= 2 \, M(v)
\end{equation}
%===============================================
is a special surface, since the expansion of the outgoing null generators vanishes on this surface and is the apparent horizon. Moreover, it is straightforward to demonstrate that $\chi^{\alpha}=\left(1,0,0,0\right)$ is null on the surface $r=r_{\rm H}$. Thus, expressing the surface element as $dS_{\alpha \beta} = \varepsilon_{\alpha \beta \mu \nu} \, e^\mu_\theta  \, e^\nu_\phi \, d \theta \, d \phi$, where $e^\mu_\theta = \delta^\mu_\theta$ and $e^\nu_\phi = \delta^\nu_\phi$ are the basis vector components on the apparent horizon, the mass within the apparent horizon is given by
%===============================================
\begin{equation}\label{HorizonMass}
M_{\rm H}=\frac{1}{8 \pi}\oint_{H} \nabla^\alpha \chi^\beta \, dS_{\alpha \beta}~.
\end{equation}
%===============================================
The above integrand turns out to be independent of the radius of the surface on which it is being evaluated, and thus, one obtains the following expression for the mass enclosed by the apparent horizon:
%===============================================
\begin{equation}\label{HorizonMassX}
M_{\rm H}= \frac{M(v)^2}{M_0}~.
\end{equation}
%===============================================
Intriguingly, the area of the apparent horizon is given by $A_{\rm H}=4\pi \, r_{\rm H}^2$ [with $r_{\rm H}$ given by Eq.~\ref{HorizonCond}], so that one may rewrite the mass within the apparent horizon as presented in Eq.~\ref{HorizonMassX} as
%===============================================
\begin{equation}\label{PseudoSmarrRelation}
M_{\rm H}=\frac{A_{\rm H}}{16 \, \pi \, M_0}=\frac{\kappa_0 \, A_{\rm H}}{4 \, \pi}~,
\quad
\kappa_{0}:= \frac{1}{4M_{0}}~,
\end{equation}
%===============================================
where $\kappa_{0}$ is the surface gravity associated with the surface $r=2M_{0}$ corresponding to the event horizon of the final black hole spacetime. The fact that $\kappa_0$ is a constant here is contradictory to the explicit formula $\kappa^{2}=-\frac{1}{2} \left(\nabla_{\mu}\chi_{\nu}\right)\left(\nabla^{\mu}\chi^{\nu}\right)$ for the surface gravity, since it depends on the gauge function $f(v)$. One may choose the gauge function, such that $f(v)=-2M(v)M'(v)$, which corresponds to the requirement that $\chi$ is null on the apparent horizon, in which case, one obtains
%===============================================
\begin{equation}\label{SurfaceGravity}
\kappa = \frac{\sqrt{1-16 M'(v)^2}}{4 M_0}~.
\end{equation}
%===============================================
Even though it appears that the surface gravity is indeed dependent on the mass function, it is straightforward to verify that if the $v$ dependence is treated as a perturbation, such that $M(v)=M_{0}+\epsilon \, \delta M(v)$, then the $v$-dependent part of $\kappa$ is a second-order term in the perturbation
%===============================================
\begin{equation}\label{SurfaceGravityPerturbed}
\kappa= \frac{1}{4M_0} - \frac{2 \delta M'(v)^2}{4 M_0} \epsilon^2 + O(\epsilon^3)~.
\end{equation}
%===============================================
Therefore, it follows that to first order, the perturbation of the surface gravity identically vanishes, and the first law $\delta M = \kappa \, \delta A / 8 \pi$ holds identically. Even then, at first sight, Eq.~\ref{SurfaceGravity} appears to be puzzling, as it seems to conflict with the expected behavior for the surface gravity, which must satisfy the exact Smarr relation $M_{\rm H} = \kappa \, A_{\rm H} /4 \pi$ (neglecting angular momentum). However, upon closer inspection, one notes that since the AKE is linear in $\chi$, the constant factor $1/M_0$ is not specified by the AKE. In an asymptotically flat spacetime, a natural choice for $M_0$ is the ADM mass $M_{\rm ADM}$, and hence, the mass enclosed by the apparent horizon may then be written as $M_{\rm H}=(M(v)/M_{\rm ADM}) \, M(v)$. This suggests that $M_{\rm H}$ can be interpreted as a rescaling of $M(v)$ by the ratio of $M(v)$ to $M_{\rm ADM}$.

To better understand this scenario, we assume that $M'(v)$ has a compact support in $v$, such that at late time, $M(v)\rightarrow M_{0}$ and $f(v)\rightarrow 0$, yielding $M_0 \rightarrow M_{\rm ADM}$, in which case one has $\chi^{\alpha}=(1,0,0,0)$. At early times, again assuming $M'(v)\rightarrow 0$ and $f(v)\rightarrow 0$, one has $\chi^{\alpha}=(M_{\rm early}/M_{\rm ADM},0,0,0)$, where $M_{\rm early}$ is the mass of the spacetime before $M'(v)$ becomes nonzero, and thus, $\chi^{\alpha}$ will differ from $(\partial/\partial t)^{\alpha}$. Since the early-time geometry of the Vaidya spacetime approximates that of a Schwarzschild black hole, then it is appropriate to rescale $\chi^{\alpha}$ by a factor $M_{\rm ADM}/M_{\rm early}$; in doing so, one obtains an early-time horizon mass consistent with the early-time ``Schwarzschild mass.'' This is an indication that the horizon mass constructed from the Komar integral for solutions of the AKE is not identical to the ``local'' (in $v$) mass of the black hole.

%=======================================================================
%-----------------------------------------------------------------------
%
%		Perturbation of GKVS
%
%-----------------------------------------------------------------------
%=======================================================================
\section{Evolution of the perturbed GKVs}\label{SGKVP}

In this section, we will consider the perturbation of a background spacetime with Killing symmetry (e.g., Schwarzschild or Kerr), which may or may not be vacuum. Since the perturbation need not respect the symmetry of the background spacetime, the perturbed spacetime does not admit an exact Killing vector field, but the perturbed spacetime will admit GKVs, as long as solutions to Eq.~\ref{AlmostKillingEquation} exist in the perturbed spacetimes. The perturbed AKE associated with the perturbation of the background spacetime will be derived in two different ways: first from a variational principle, where the perturbation of the action presented in Eq.~\ref{ActionComp} will be considered, and then from the direct perturbation of the AKE itself. We will verify that the results arising out of these two different approaches match.

%===============================================
%-----------------------------------------------
%
%		NOTATION
%
%-----------------------------------------------
%===============================================
\subsection{Notations and conventions} \label{SSNotConv}

Before proceeding further, it is perhaps appropriate to settle the notations and the conventions that we will use for this section and the remainder of this article. With the exception of the GKVs, often denoted as $\chi^\mu$, ``barred'' symbols will be used to denote exact quantities; for instance $\bar{g}_{\alpha \beta}$, $\bar{\nabla}_\alpha$, and $\bar{R}_{\alpha \beta}$, respectively, denote the exact metric, connection, and Ricci tensor. Unbarred geometric quantities will be used to denote background quantities. Metric perturbations will be denoted $h_{\alpha \beta}$, and are defined by
%===============================================
\begin{align}\label{metric_pert}
h_{\alpha \beta} := \bar{g}_{\alpha \beta} - g_{\alpha \beta}~.
\end{align}
%===============================================
\noindent Indices are raised and lowered according to the background metric $g_{\alpha \beta}$. Killing vector fields for the background spacetime will be denoted $\xi^{\mu}$, and we define $\delta \xi^\mu$ to be the difference between the GKV and the exact Killing vector, in the following manner:
%===============================================
\begin{align}\label{chi_pert}
\delta \xi^\alpha := \chi^\alpha - \xi^\alpha~.
\end{align}
%===============================================
In general, the prefix $\delta$ will denote differences between the exact and background quantities (which we will later assume to be small compared to background values), and the prefix $\Delta$ will denote first-order variations.

%===============================================
%-----------------------------------------------
%
%		Perturbation of AKE Action
%
%-----------------------------------------------
%===============================================
\subsection{Perturbation of the action principle yielding the almost Killing equation}

As emphasized before, we assume that the background metric $g_{\mu \nu}$ admits a Killing vector field $\xi^\mu$. Since $h_{\mu \nu}$ is a perturbation over and above the background spacetime, it is legitimate to assume that $h_{\mu \nu} \ll g_{\mu \nu}$, and for some normalization of the background Killing vector field $\xi^\mu$, we also assume $\delta \xi^\mu \ll \xi^\mu$
\footnote{Technically speaking, this is achieved by introducing two parameters, $\epsilon_{1}$ and $\epsilon_{2}$, with $\epsilon_{1,2} \ll 1$ and then expanding the metric as $\bar{g}_{\mu \nu}=g_{\mu \nu}+\epsilon_{1} h_{\mu \nu}$ and the GKV as $\chi^{\alpha}=\xi^{\alpha}+\epsilon_{2} \delta \xi^{\alpha}$. Then keeping terms linear in $\epsilon_{1}$ and $\epsilon_{2}$ will provide first-order perturbation, while terms $\mathcal{O}(\epsilon_{1,2}^{2})$ yield the second-order perturbations.}.
The action presented in Eq.~\ref{ActionComp} may then be expanded in $h_{\mu \nu}$, $\delta \xi^{\mu}$, and $\delta \mu:=\bar{\mu}-\mu$, keeping terms up to quadratic order in each. We treat the expansions in $h_{\mu \nu}$ and $\delta \xi^{\mu}$ independently, so that we keep terms of the form $h_{\cdot \cdot }h_{\cdot \cdot} \delta \xi^{\cdot} \delta \xi^{\cdot}$. For simplicity, we assume the quantity $\delta \mu$ to be independent of the spacetime coordinates. It should be emphasized that $\delta \xi_{\mu} \neq \bar{g}_{\mu \nu}(\chi^\nu - \xi^\nu)$, since indices are raised and lowered with the background metric. We begin by writing down the Lagrangian from the action functional presented in Eq.~\ref{ActionComp}, which has the following explicit form in the spacetime with metric $\bar{g}_{\mu \nu}$ :
%===============================================
\begin{align}\label{Lagrangian_01}
\bar{L}=-\frac{1}{4}\left(\bar{g}_{\nu \alpha}\bar{\nabla}_{\mu}\chi^{\alpha}+\bar{g}_{\mu \alpha}\bar{\nabla}_{\nu}\chi^{\alpha}\right)\left(\bar{g}^{\mu \beta}\bar{\nabla}_{\beta}\chi^{\nu}+\bar{g}^{\nu \beta}\bar{\nabla}_{\beta}\chi^{\mu}\right)+\frac{\bar{\mu}}{2}\left(\bar{\nabla}_{\mu}\chi^{\mu}\right)^{2}~.
\end{align}
%===============================================
We wish to express the Lagrangian presented above solely in terms of the background metric $g_{\mu \nu}$, the perturbation $h_{\mu \nu}$ and the perturbation $\delta \xi^{\mu}$ along with $\delta \mu$. Since the Lagrangian depicted in Eq.~\ref{Lagrangian_01} consists of terms quadratic in $\chi^\mu$ and its derivatives, it can be rewritten in the following form:
%===============================================
\begin{align}\label{Lagrangian_form}
        \bar{L}
        =
        \chi^\alpha \, \chi^\beta \,
		L_{\alpha \beta}
		+
		\chi^\mu \, \nabla^\alpha \chi^{\beta} \,
		L_{\mu \alpha \beta}
		+
		\nabla^\nu \chi^{\mu} \, \nabla^\beta \chi^{\alpha} \,
		L_{\mu \nu \alpha \beta}~,
\end{align}
%===============================================
where the tensors $L_{\alpha \beta}$, $L_{\mu \alpha \beta}$, and $L_{\mu \nu \alpha \beta}$ depend on the background metric $g_{\mu \nu}$, the perturbation $h_{\mu \nu}$, the derivatives of $h_{\mu \nu}$, and $\bar{\mu}$. Note that in writing Eq.~\ref{Lagrangian_form}, we make no assumption about $h_{\mu \nu}$; Eq.~\ref{Lagrangian_form} should hold to all orders in the metric perturbation $h_{\mu \nu}$. Using the above decomposition of the Lagrangian, the action can similarly be written down in the following form:
%===============================================
\begin{align}\label{Action_form}
\mathcal{A}[\delta \xi^\cdot]
    &=\int_{\mathcal{V}} d^{4}x\sqrt{-g}
        \Bigg[
        \chi^\alpha \, \chi^\beta \,
		M_{\alpha \beta}
		+
		\chi^\mu \, \nabla^\alpha \chi^{\beta} \,
		M_{\mu \alpha \beta}
		+
		\nabla^\nu \chi^{\mu} \, \nabla^\beta \chi^{\alpha} \,
		M_{\mu \nu \alpha \beta}
        \Bigg]~,
\end{align}
%===============================================
where we define
%===============================================
\begin{align}\label{Action_coeffs}
M_{\sigma_1 \sigma_2 ...}:=\sqrt{\bar{g}/g} \, L_{\sigma_1 \sigma_2 ...}~.
\end{align}
%===============================================
One may at this point perform the variation of the action with respect to $\delta \xi^{\alpha}$, the perturbed GKV, without explicit knowledge of the tensors $M_{\sigma_1 \sigma_2 ...}$. Up to boundary terms, the first-order variation of the action takes the following form:
%===============================================
\begin{equation} \label{PAKE-ActionVariation}
	\begin{aligned}
	\Delta \mathcal{A}
	= \int d^4x \sqrt{-g} \,
	\Delta \delta \xi^\sigma
	\mathbb{E}_\alpha~,
	\end{aligned}
\end{equation}
%===============================================
where
%===============================================
\begin{equation} \label{PAKE-ActionVariationEoML}
	\begin{aligned}
	\mathbb{E}_\alpha
	&
	:=
    \frac{1}{2} \biggl\{
    2 \,
    \nabla{^\beta}\chi{^\nu}
    \biggl[
      \frac{1}{2}
      (
      L_{\beta \nu \gamma \alpha}
      +
      L_{\gamma \alpha \beta \nu}
      )
      (
      h^{\sigma \tau} \nabla{^\gamma} h_{\sigma \tau}
      -
      \nabla{^\gamma}h
      )
      -
      (
        \nabla{^\gamma} L_{\beta \nu \gamma \alpha}
        +
        \nabla{^\gamma} L_{\gamma \alpha \beta \nu}
      )
      +
      L_{\alpha \beta \mu}
      -
      L_{\nu \beta \alpha}
    \biggr] \\
    & \qquad \quad
    +
    2 \,
    \chi{^\nu}
    \biggl[
      \frac{1}{2}  L_{\nu \beta \alpha}
      (
        h^{\gamma \sigma} \nabla{^\beta} h_{\gamma \sigma}
        -
        \nabla{^\beta}h
      )
      -
      \nabla{^\beta} L_{\nu \beta \alpha}
      +
      L_{\alpha \nu}
      +
      L_{\nu \alpha}
    \biggr]
    -
    2 \,
    \nabla{^\gamma}\nabla{^\beta}\chi{^\nu}
    (
      L_{\beta \nu \gamma \alpha}
      +
      L_{\gamma \alpha \beta \nu}
    )
    \biggr\}~.
	\end{aligned}
\end{equation}
%===============================================
Keeping in mind $\chi^\alpha=\xi^\alpha + \delta \xi^\alpha$, the perturbed AKE may then be written as
%===============================================
\begin{equation} \label{PAKE-ActionVariationEoM}
\frac{\bar{g}^{\mu \alpha} \, \mathbb{E}_\alpha}{\sqrt{\bar{g}/g}} = 0~.
\end{equation}
%===============================================
The factor of $1/\sqrt{\bar{g}/g}$ is included because one typically factors out the volume element from the functional derivative when deriving field equations [as was done when going from Eq.~\ref{pert_ake_var} to Eq.~\ref{pert_ake} in the Appendix]. Again, we emphasize that the analysis presented here does not require that $h_{\mu \nu}$ be small; the result in Eq.~\ref{PAKE-ActionVariationEoM} holds to all orders in $h_{\mu \nu}$. Thus, one may expand the tensors $L_{\sigma_1 \sigma_2 ...}$ in various powers of the gravitational perturbation in the following manner:
%===============================================
\begin{equation} \label{LagPerturbedTensorsDecomp}
	\begin{aligned}
		L_{\alpha \beta}
		= &
		L^{0}_{\alpha \beta} + L^{1}_{\alpha \beta} + L^{2}_{\alpha \beta} + O(h_{\cdot\cdot}^3),
    \\
		L_{\mu \alpha \beta}
		= &
		L^{0}_{\mu \alpha \beta} + L^{1}_{\mu \alpha \beta} + L^{2}_{\mu \alpha \beta} + O(h_{\cdot\cdot}^3),
		\\
		L_{\mu \nu \alpha \beta}
		= &
		L^{0}_{\mu \nu \alpha \beta} + L^{1}_{\mu \nu \alpha \beta} + L^{2}_{\mu \nu \alpha \beta} + O(h_{\cdot\cdot}^3) ~.
	\end{aligned}
\end{equation}
%===============================================
It turns out that to zeroth order, one has $L^{0}_{\alpha \beta}=0$, $L^{0}_{\mu \alpha \beta}=0$, while
%===============================================
\begin{equation} \label{PAKE-ActionPerturbedTensors0}
	L^{0}_{\mu \nu \alpha \beta}
	=
	\frac{1}{2}
	\left(
	\bar{\mu}~ g_{\alpha \beta} g_{\mu \nu}
	-
	g_{\alpha \nu} g_{\beta \mu}
	-
	g_{\alpha \mu} g_{\beta \nu}
	\right)~.
\end{equation}
%===============================================
To first order in $h_{\mu \nu}$, the tensors $L^{1}_{\alpha \beta}$, $L^{1}_{\mu \alpha \beta}$, and $L^{1}_{\mu \nu \alpha \beta}$ can be expressed as linear functions of the gravitational perturbation $h_{\mu \nu}$ as
%===============================================
\begin{equation} \label{PAKE-ActionPerturbedTensors1}
	\begin{aligned}
		L^{1}_{\alpha \beta}
		=
		0~,
		\quad
		L^{1}_{\mu \alpha \beta}
		=
		\frac{1}{2} \biggl\{
		\bar{\mu} g_{\alpha \beta} \nabla_\mu h - 2 \nabla_\mu h_{\alpha \beta}
		\biggr\}~,
		\quad
		L^{1}_{\mu \nu \alpha \beta}
		= &
		\frac{1}{2} \biggl\{
	    h_{\alpha \mu} g_{\beta \nu} - g_{\alpha \mu} h_{\beta \nu}
		\biggr\}~ .
	\end{aligned}
\end{equation}
%===============================================
Finally, we present the second-order terms in the perturbation $h_{\mu \nu}$ as
%===============================================
\begin{equation} \label{PAKE-ActionPerturbedTensors2}
	\begin{aligned}
		L^{2}_{\alpha \beta}
		= &
		\frac{1}{8}
		\biggl\{
		\bar{\mu} \nabla_{\alpha}h \nabla_{\beta}h
		-
		2 \nabla_{\alpha} h^{\sigma \tau} \nabla_{\beta} h_{\sigma \tau}
		\biggr\},
		\\
		L^{2}_{\mu \alpha \beta}
		= &
		\frac{1}{2}
		\biggl\{
		2 h_{\alpha}{^\sigma} \nabla_{\mu} h_{\beta \sigma}
		-
		\bar{\mu} g_{\alpha \beta} h^{\sigma \tau} \nabla_{\mu} h_{\sigma \tau}
		\biggr\},
		\\
		L^{2}_{\mu \nu \alpha \beta}
		= &
		\frac{1}{2}
		\biggl\{
		h_{\alpha \mu} h_{\beta \nu}
		-
		g_{\beta \nu} h_{\alpha}{^\sigma} h_{\mu \sigma}
		\biggr\}~.
	\end{aligned}
\end{equation}
%===============================================
In what follows, we consider in detail the expansion of the Lagrangian to first order in the metric perturbation $h_{\mu \nu}$, and as we shall demonstrate, the resulting perturbed AKE is consistent with the expansion coefficients determined above.

%=======================================================================
%=======================================================================

%===============================================
%-----------------------------------------------
%
%		Explicit Perturbation of Action (O(h))
%
%-----------------------------------------------
%===============================================
\subsection{Explicit perturbation of the action to first order in the metric}

The expression for the AKE given in Eq.~\ref{PAKE-ActionVariationEoML} is rather complicated and somewhat opaque; it is perhaps more illustrative to show explicitly that the derivation of the perturbed AKE to first order in the metric perturbations $h_{\mu \nu}$, simplifying the expressions along the way. To obtain the expansion of the Lagrangian to first order in $h_{\mu \nu}$, one may use Eqs. \ref{PAKE-ActionPerturbedTensors0} and \ref{PAKE-ActionPerturbedTensors1}
%-unc or the identities presented in \ref{AppB}
to obtain the following Lagrangian:
%===============================================
\begin{align}\label{lag_density_02}
\bar{L}&=-\frac{1}{4}\left[\left(g_{\nu \alpha}+h_{\nu \alpha}\right)\nabla_{\mu}\chi^{\alpha}
+\left(g_{\mu \alpha}+h_{\mu \alpha}\right)\nabla_{\nu}\chi^{\alpha}+\chi^{\rho}\nabla_{\rho}h_{\mu \nu}\right]
\nonumber
\\
&\hskip 2 cm \times
\left[\left(g^{\mu \beta}-h^{\mu \beta}\right)\nabla_{\beta}\chi^{\nu}
+\left(g^{\nu \beta}-h^{\nu \beta}\right)\nabla_{\beta}\chi^{\mu}+\chi^{\sigma}\nabla_{\sigma}h^{\mu \nu}\right]
+\frac{\bar{\mu}}{2}\Big[\left(\nabla_{\mu}\chi^{\mu}\right)^{2}+\chi^{\alpha}\nabla_{\alpha}h\left(\nabla_{\mu}\chi^{\mu}\right)\Big]~.
\end{align}
%===============================================
Having expressed the Lagrangian explicitly in terms of the background metric $g_{\mu \nu}$ and the perturbation $h_{\alpha \beta}$, we now expand the GKV field in terms of the background Killing field $\xi^{\mu}$ and the perturbation $\delta \xi^{\mu}$. One can see that Eq.~\ref{lag_density_02} has the form of Eq.~\ref{Lagrangian_form}, and it is not difficult to verify that the Lagrangians are equivalent for the coefficients given in Eqs. ~\ref{LagPerturbedTensorsDecomp}--\ref{PAKE-ActionPerturbedTensors1}. Using the Killing equation for $\xi^{\mu}$, i.e., setting $\nabla_{\mu}\xi_{\nu}+\nabla_{\nu}\xi_{\mu}=0$, the Lagrangian presented in Eq.~\ref{lag_density_02} can be further simplified. In particular, it is worth emphasizing that the on-shell value of the action for the background Killing vector field identically vanishes, and thus, the Lagrangian density given in Eq.~\ref{lag_density_02} becomes
%===============================================
\begin{align}\label{lag_action_first}
\bar{L}&=-\frac{1}{2}\Big[g_{\nu \alpha}\nabla_{\mu}\delta \xi^{\alpha}+h_{\nu \alpha}\nabla_{\mu}\xi^{\alpha}+h_{\nu \alpha}\nabla_{\mu}\delta \xi^{\alpha}+g_{\mu \alpha}\nabla_{\nu}\delta \xi^{\alpha}+h_{\mu \alpha}\nabla_{\nu}\xi^{\alpha}+h_{\mu \alpha}\nabla_{\nu}\delta \xi^{\alpha}+\left(\xi^{\rho}+\delta \xi^{\rho}\right)\nabla_{\rho}h_{\mu \nu}\Big]
\nonumber
\\
&\hskip 0.5 cm \times\Big[\nabla^{\mu}\delta \xi^{\nu}-h^{\mu \beta}\nabla_{\beta}\xi^{\nu}-h^{\mu \beta}\nabla_{\beta}\delta \xi^{\nu}+\frac{1}{2}\left(\xi^{\sigma}+\delta \xi^{\sigma}\right)\nabla_{\sigma}h^{\mu \nu}\Big]+\frac{\bar{\mu}}{2}\Big[\left(\nabla_{\mu}\delta\xi^{\mu}\right)^{2}+\left(\xi^{\alpha}+\delta \xi^{\alpha}\right)\nabla_{\alpha}h\left(\nabla_{\mu}\delta \xi^{\mu}\right)\Big]+\mathcal{O}(h_{\cdot \cdot}^2)~.
\end{align}
%===============================================
Even though the above Lagrangian density looks sufficiently complicated, we can reduce it to a very simple form by dividing the above into three categories: (a) terms quadratic in the derivatives of $\delta \xi^{\alpha}$, (b) terms linear in the derivatives of $\delta \xi^{\alpha}$, and (c) terms independent of derivatives of $\delta \xi^{\alpha}$. The terms quadratic in the derivative of $\delta \xi^{\alpha}$ yield
%-unc (see \ref{AppB})
%===============================================
\begin{align}\label{quad_terms}
\textrm{Quadratic~terms}=\frac{1}{2}\left[\left(\mu+\delta \mu\right)\delta^{\mu}_{\alpha}\delta ^{\nu}_{\beta}-\left(g_{\alpha \beta}+h_{\alpha \beta}\right)\left(g^{\mu \nu}-h^{\mu \nu}\right)-\left(\delta^{\nu}_{\alpha}+h^{\nu}_{\alpha}\right)\left(\delta ^{\mu}_{\beta}-h^{\mu}_{\beta}\right)\right]\left(\nabla_{\mu}\delta \xi^{\alpha}\nabla_{\nu}\delta \xi^{\beta}\right)~,
\end{align}
%===============================================
while the terms linear in the derivative of $\delta \xi^{\alpha}$ become
%-unc (see \ref{AppB})
%===============================================
\begin{align}
\textrm{Linear~terms}&=\frac{\bar{\mu}}{2}\Big[\left(\xi^{\alpha}+\delta \xi^{\alpha}\right)\nabla_{\alpha}h\left(\nabla_{\mu}\delta \xi^{\mu}\right)\Big]
-\frac{1}{2}\Big[g_{\nu \alpha}\nabla_{\mu}\delta \xi^{\alpha}+g_{\mu \alpha}\nabla_{\nu}\delta \xi^{\alpha}\Big]\times \Big[-h^{\mu \beta}\nabla_{\beta}\xi^{\nu}+\frac{1}{2}\left(\xi^{\sigma}+\delta \xi^{\sigma}\right)\nabla_{\sigma}h^{\mu \nu}\Big]
\nonumber
\\
&\hskip 1 cm -\frac{1}{2}\nabla^{\mu}\delta \xi^{\nu}\Big[h_{\nu \alpha}\nabla_{\mu}\xi^{\alpha}+h_{\mu \alpha}\nabla_{\nu}\xi^{\alpha}+\left(\xi^{\rho}+\delta \xi^{\rho}\right)\nabla_{\rho}h_{\mu \nu}\Big]~.
\end{align}
%===============================================
Finally, as indicated in the expression for $L_{\alpha \beta}$ in Eq.~\ref{PAKE-ActionPerturbedTensors1} the terms involving no derivatives of $\delta \xi^{\alpha}$ is $\mathcal{O}(h_{\cdot \cdot}^{2})$, and hence, will not contribute in our subsequent discussion regarding the determination of the action functional of the perturbed AKE. Thus we have computed the Lagrangian of the GKV field involving linear order terms in the perturbation $h_{\mu \nu}$ and up to quadratic order terms in the perturbed GKV $\delta \xi^{\mu}$. However, computation of the action functional requires multiplication of the above Lagrangian by a factor of $\sqrt{-\bar{g}}$, where $\bar{g}$ is the determinant of the perturbed metric $\bar{g}_{\alpha \beta}$. Therefore, the complete action for the perturbed Killing vector field $\delta \xi^{\mu}$ takes the following form:
%===============================================
\begin{align}\label{lagrangian_exp}
\mathcal{A}[\delta \xi^\cdot]&=\int_{\mathcal{V}} d^{4}x\sqrt{-g} \Bigg\{\Bigg[\frac{1}{2}\left(\mu~ \delta^{\mu}_{\alpha}\delta ^{\nu}_{\beta}-g_{\alpha \beta}g^{\mu \nu}-\delta^{\nu}_{\alpha}\delta ^{\mu}_{\beta}\right)+\frac{1}{4}h\left(\mu~ \delta^{\mu}_{\alpha}\delta ^{\nu}_{\beta}-g_{\alpha \beta}g^{\mu \nu}-\delta^{\nu}_{\alpha}\delta ^{\mu}_{\beta}\right)
\nonumber
\\
&\hskip 1 cm +\frac{1}{2}\left(1+\frac{1}{2}h\right)\delta \mu~\delta^{\mu}_{\alpha}\delta ^{\nu}_{\beta}+\frac{1}{2}\left(g_{\alpha \beta}h^{\mu \nu}-h_{\alpha \beta}g^{\mu \nu}+\delta^{\nu}_{\alpha}h^{\mu}_{\beta}-h^{\nu}_{\alpha}\delta ^{\mu}_{\beta}\right)\Bigg]
\left(\nabla_{\mu}\delta \xi^{\alpha}\nabla_{\nu}\delta \xi^{\beta}\right)
\nonumber
\\
&\hskip 1 cm +\Bigg[\frac{\bar{\mu}}{2}\Big[\left(\xi^{\alpha}+\delta \xi^{\alpha}\right)\nabla_{\alpha}h\left(\nabla_{\mu}\delta \xi^{\mu}\right)\Big]
-\Big[\left(\nabla_{\mu}\delta \xi^{\alpha}\right)h_{\alpha \beta}\nabla^{\mu}\xi^{\beta}+\left(\nabla_{\mu}\delta \xi^{\alpha}\right)h^{\mu \beta}\nabla_{\alpha}\xi_{\beta}\Big]
\nonumber
\\
&\hskip 1.5 cm -g_{\nu \alpha}\nabla_{\mu}\delta \xi^{\alpha}\left(\xi^{\sigma}+\delta \xi^{\sigma}\right)\nabla_{\sigma}h^{\mu \nu}\Bigg]\Bigg\}~.
\end{align}
%===============================================
Having derived the action to linear order in the gravitational perturbation $h_{\mu \nu}$, we can determine an arbitrary variation of the action for variation of the perturbed Killing vector field $\delta \xi^{\mu}$, which when set to zero should yield the corresponding perturbed AKE. The
%-unc above variation of the action has been carried out in \ref{AppB} and the
final expression for the variation, ignoring any boundary contribution, takes the following form:
%===============================================
\begin{align}\label{Eom_variation}
\Delta \mathcal{A}&=\int _{\mathcal{V}}d^{4}x~\sqrt{-g}~\delta \left(\delta \xi^{\alpha}\right)\left(1+\frac{1}{2}h\right)\left(g_{\alpha \beta}+h_{\alpha \beta}\right)\Bigg\{\Big[\left(1-\mu\right)\nabla^{\beta}\left(\nabla_{\nu}\delta \xi^{\nu}\right)+\square \delta \xi^{\beta}+R{^\beta}_{\rho}\delta \xi^{\rho}\Big]
\nonumber
\\
&\hskip 0.5 cm -\delta \mu~\nabla^{\beta}\nabla_{\sigma}\delta \xi^{\sigma}-h^{\mu \nu}\nabla_{\mu}\nabla_{\nu}\delta \xi^{\beta}
+2\left(\nabla^{(\mu}\delta \xi^{\nu)}\right)\left(\nabla_{\mu}h^{\beta}_{\nu}-\frac{1}{2}\nabla^{\beta}h_{\mu \nu}\right)
-R{^\beta}_{\rho \sigma \mu}h^{\mu \rho}\delta \xi^{\sigma}
\nonumber
\\
&\hskip 0.5 cm +\left(\frac{1-\mu-\delta \mu}{2}\right)\nabla^{\beta}\left[\left(\xi^{\rho}+\delta \xi^{\rho}\right)\nabla_{\rho}h \right]-h^{\beta \mu}\left(1-\mu-\delta \mu\right)\nabla_{\mu}\left(\nabla_{\nu}\delta \xi^{\nu}\right)\Bigg\}~,
\end{align}
%===============================================
where we have neglected all the terms quadratic in the gravitational perturbation $h_{\mu \nu}$. Setting the variation $\Delta \mathcal{A}$ to zero, for arbitrary variation of the perturbation of the Killing vector field $\delta \xi^{\mu}$, we obtain the following dynamical equation for the perturbed GKV field $\delta \xi^{\mu}$:
%===============================================
\begin{align}\label{pake_action}
\big(1&-\mu\big)\nabla^{\beta}\left(\nabla_{\nu}\delta \xi^{\nu}\right)+\square \delta \xi^{\beta}+R{^\beta}_{\rho}\delta \xi^{\rho}=J^{\beta}~,
\\
J^{\beta}&=\delta \mu~\nabla^{\beta}\nabla_{\sigma}\delta \xi^{\sigma}+h^{\mu \nu}\nabla_{\mu}\nabla_{\nu}\delta \xi^{\beta}
-2\left(\nabla^{(\mu}\delta \xi^{\nu)}\right)\left(\nabla_{\mu}h^{\beta}_{\nu}-\frac{1}{2}\nabla^{\beta}h_{\mu \nu}\right)
+R{^\beta}_{\rho \sigma \mu}h^{\mu \rho}\delta \xi^{\sigma}
\nonumber
\\
&\hskip 0.5 cm -\left(\frac{1-\mu-\delta \mu}{2}\right)\nabla^{\beta}\left[\left(\xi^{\rho}+\delta \xi^{\rho}\right)\nabla_{\rho}h \right]+h^{\beta \mu}\left(1-\mu-\delta \mu\right)\nabla_{\mu}\left(\nabla_{\nu}\delta \xi^{\nu}\right)~.
\end{align}
%===============================================
The above provides the dynamical equation for the perturbed Killing vector field $\delta \xi^{\alpha}$ arising from the variation of the action. One can verify that this expression is equivalent to that obtained from Eq.~\ref{PAKE-ActionVariationEoML}; we have done this using the xAct package for \textit{Mathematica}. In the subsequent discussion, we will discuss explicitly the derivation of this equation from the perturbation of the AKE itself to first order in the metric perturbations. This will depict the internal consistency of the results derived in this work.

%===============================================
%-----------------------------------------------
%
%		Perturbation of Action
%
%-----------------------------------------------
%===============================================
\subsection{Perturbation of the almost Killing equation to first order in the metric}

We have derived the evolution equation for the perturbed GKV field to first order in the metric perturbations, starting from the variation of the perturbed action for the GKV field. As we will show in this section, the same equation can also be derived from direct perturbation of the AKE itself. As before, we assume that the background metric $g_{\mu \nu}$ admits a Killing vector field $\xi^\mu$ and also that $h_{\mu \nu} \ll g_{\mu \nu}$ and $\delta \xi^\mu \ll \xi^\mu$ (see also footnote 2). Thus, we will expand the AKE given in Eq.~\ref{AlmostKillingEquation} in the perturbed spacetime with metric $\bar{g}_{\mu \nu}$ to first order in $\delta \xi^\mu$ and $h_{\mu \nu}$ each, again assuming that the expansions are independent (so that we keep terms of the form $h_{\cdot \cdot} \delta \xi^{\cdot}$). It is convenient to first present the expansion of the following geometric quantities to linear order in the gravitational perturbation $h_{\mu \nu}$
%-unc (for the derivation of these identities, see \ref{AppC})
:
%===============================================
\begin{align}\label{pert_ricci}
\delta R_{\mu \nu}=\frac{1}{2}\left(-\square h_{\mu \nu}-\nabla_{\mu}\nabla_{\nu}h+\nabla_{\mu}\nabla_{\alpha}h^{\alpha}_{\nu}+\nabla_{\nu}\nabla_{\alpha}h^{\alpha}_{\mu}+R_{\beta \mu}h^{\beta}_{\nu}+R_{\beta \nu}h^{\beta}_{\mu}-2R_{\alpha \mu \beta \nu}h^{\alpha \beta}\right)~,
\end{align}
%===============================================
%===============================================
\begin{align}\label{identity_1}
\bar{\nabla}_{\alpha}\bar{\nabla}_{\beta}V^{\beta}&=\delta _{\mu}^{\beta}\bar{\nabla}_{\alpha}\bar{\nabla}_{\beta}V^{\mu}
=\nabla_{\alpha}\left(\nabla_{\beta}V^{\beta}\right)+\nabla_{\alpha}\left(\delta \Gamma^{\beta}_{\beta \rho}V^{\rho}\right)+\delta \Gamma^{\beta}_{\alpha \rho}\nabla_{\beta}V^{\rho}-\delta\Gamma^{\rho}_{\alpha \beta}\nabla_{\rho}V^{\beta}
\nonumber
\\
&=\nabla_{\alpha}\left(\nabla_{\beta}V^{\beta}\right)+\nabla_{\alpha}\left(\delta \Gamma^{\beta}_{\beta \rho}V^{\rho}\right)~,
\end{align}
%===============================================
%===============================================
\begin{align}\label{identity_square}
\bar{\square}V^{\mu}&=\left(g^{\alpha \beta}-h^{\alpha \beta}\right)\nabla_{\alpha}\nabla_{\beta}V^{\mu}+\left(-\frac{1}{2}\nabla^{\mu}h_{\alpha \rho}+\nabla_{\alpha}h^{\mu}_{\rho}\right)\left(\nabla^{\alpha}V^{\rho}+\nabla^{\rho}V^{\alpha}\right)-\nabla_{\rho}V^{\mu}\left(\nabla_{\alpha}h^{\alpha \rho}-\frac{1}{2}\nabla^{\rho}h\right)
\nonumber
\\
&\hskip 1 cm +\frac{1}{2}V^{\rho}\left(\square h^{\mu}_{\rho}+\nabla_{\rho}\nabla_{\beta}h^{\beta\mu}-\nabla^{\mu}\nabla^{\beta}h_{\beta \rho}+R_{\sigma \rho}h^{\sigma \mu}-R^{\sigma \mu}h_{\sigma \rho}\right)~,
\end{align}
%===============================================
where $V^{\mu}$ is an arbitrary vector field. In deriving the above identities, we have used various properties of the Riemann tensor, e.g., $R_{\alpha \beta \mu \nu}=R_{\nu \mu \beta \alpha}$ among others. Applying all these identities to the AKE in the perturbed spacetime and imposing the Lorenz gauge condition $\nabla_{\alpha}h^{\alpha}_{\rho}=(1/2)\nabla_{\rho}h$, we obtain
%===============================================
\begin{align}
\Big(g^{\alpha \beta}&-h^{\alpha \beta}\Big)\nabla_{\alpha}\nabla_{\beta}\chi^{\mu}+\left(-\frac{1}{2}\nabla^{\mu}h_{\alpha \rho}+\nabla_{\alpha}h^{\mu}_{\rho}\right)\left(\nabla^{\alpha}\chi^{\rho}+\nabla^{\rho}\chi^{\alpha}\right)+R{^\mu}_{\beta}\chi^{\beta}
\nonumber
\\
&\hskip 1 cm +\left(1-\bar{\mu}\right)\bar{g}^{\mu \alpha}\left[\nabla_{\alpha}\left(\nabla_{\beta}\chi^{\beta}\right)+\frac{1}{2}\nabla_{\alpha}\left(\chi^{\rho}\nabla_{\rho}h\right)\right]-R^{\mu}_{~\alpha \beta \sigma}h^{\alpha \sigma}\chi^{\beta}
\nonumber
\\
&\hskip 3.5 cm +\left\{\nabla_{\alpha}\left(1-\bar{\mu}\right)\right\}\bar{g}^{\mu \alpha}\left[\left(\nabla_{\beta}\chi^{\beta}\right)+\frac{1}{2}\left(\chi^{\rho}\nabla_{\rho}h\right)\right]=0~,
\end{align}
%===============================================
which is valid up to linear order in the gravitational perturbation $h_{\mu \nu}$.

At this point, we have not yet expanded in the GKV field $\chi^\mu$; we do this now. We make use of the wave equation for the background Killing vector field $\xi^\mu$ given in Eq.~\ref{DivKillingEquation} and other properties of Killing vectors to obtain
%===============================================
\begin{align}
\Big(g^{\alpha \beta}&-h^{\alpha \beta}\Big)\nabla_{\alpha}\nabla_{\beta}\delta \xi^{\mu}+\left(-\frac{1}{2}\nabla^{\mu}h_{\alpha \rho}+\nabla_{\alpha}h^{\mu}_{\rho}\right)\left(\nabla^{\alpha}\delta \xi^{\rho}+\nabla^{\rho}\delta \xi^{\alpha}\right)+R{^\mu}_{\beta}\delta \xi^{\beta}
\nonumber
\\
&\hskip 1 cm +\left(1-\bar{\mu}\right)\bar{g}^{\mu \alpha}\left\{\nabla_{\alpha}\left(\nabla_{\beta}\delta \xi^{\beta}\right)+\frac{1}{2}\nabla_{\alpha}\left[\left(\xi^{\rho}+\delta \xi^{\rho}\right)\nabla_{\rho}h\right]\right\}-R^{\mu}_{~\alpha \beta \sigma}h^{\alpha \sigma}\delta \xi^{\beta}
\nonumber
\\
&\hskip 4 cm +\left\{\nabla_{\alpha}\left(1-\bar{\mu}\right)\right\}\bar{g}^{\mu \alpha}\left[\left(\nabla_{\beta}\delta \xi^{\beta}\right)+\frac{1}{2}\left(\xi^{\rho}\nabla_{\rho}h+\delta\xi^{\rho}\nabla_{\rho}h\right)\right]=0~.
\end{align}
%===============================================
This is our result for the perturbed AKE. The above equation has been derived under very general conditions, without any assumptions about the nature of the perturbation. Thus, it is possible to express the above equation in several different ways, under different assumptions, which we will list below. First, we rewrite the above evolution equation for the perturbed GKV field $\delta \xi^{\mu}$ in the following form:
%===============================================
\begin{align}
&\square \delta \xi^{\mu}+R{^\mu}_{\nu}\delta \xi^{\nu}+g^{\mu \alpha}\nabla_{\alpha}\left\{\left(1-\mu\right)\left(\nabla_{\sigma}\delta \xi^{\sigma}\right)\right\}=j^{\mu}~,\label{pake_form_02}
\\
&j^{\mu}=h^{\alpha \beta}\left[\nabla_{\alpha}\nabla_{\beta}\delta \xi^{\mu}+R^{\mu}_{~\alpha \rho \beta}\delta \xi^{\rho}+\delta^{\mu}_{\alpha}\nabla_{\beta}\left\{\left(1-\mu-\delta \mu\right)\left(\nabla_{\sigma}\delta \xi^{\sigma}\right)\right\}\right]+g^{\mu \alpha}\nabla_{\alpha}\left(\delta \mu~ \nabla_{\sigma}\delta \xi^{\sigma}\right)
\nonumber
\\
&\hskip 1 cm -2\left(\nabla_{\alpha}h^{\mu}_{\rho}-\frac{1}{2}\nabla^{\mu}h_{\alpha \rho}\right)\nabla^{(\alpha}\delta \xi^{\rho)}
-\frac{1}{2}\left(1-\mu-\delta \mu\right)g^{\mu \alpha}\nabla_{\alpha}\left[\left(\xi^{\rho}+\delta \xi^{\rho}\right)\nabla_{\rho}h\right]
\nonumber
\\
&\hskip 4 cm -\frac{1}{2}\nabla_{\alpha}\left(1-\mu-\delta \mu\right)g^{\mu \alpha}\left[\left(\xi^{\rho}+\delta \xi^{\rho}\right)\nabla_{\rho}h\right]~.
\end{align}
%===============================================
Upon comparison, we find that this evolution equation for the perturbed GKV $\delta \xi^{\mu}$ is identical to what we had derived from the action, i.e., to Eq.~\ref{pake_action}, except for the terms involving derivatives of $\mu$ and $\delta \mu$, respectively. This is because, while deriving Eq.~\ref{pake_action}, we have assumed for convenience that $\mu$ and $\delta \mu$ are constants, while that is not the case for the derivation presented above. If we assume that $\mu$ for the background spacetime is constant, and $\delta \mu$ to be a scalar function then the dynamics of the perturbed GKV is determined by
%===============================================
\begin{align}\label{pake_form_04}
\big(g^{\alpha \beta}&-h^{\alpha \beta}\big)\left[\nabla_{\alpha}\nabla_{\beta}\delta \xi^{\mu}+R^{\mu}_{~\alpha \rho \beta}\delta \xi^{\rho}\right]=j^{\mu}~,
\\
j^{\mu}&=-\left(1-\mu-\delta \mu\right)\left[\left(g^{\mu \alpha}-h^{\mu \alpha}\right)\nabla_{\alpha}\left(\nabla_{\sigma}\delta \xi^{\sigma}\right)
+\frac{1}{2}g^{\mu \alpha}\nabla_{\alpha}\left\{\left(\xi^{\rho}+\delta \xi^{\rho}\right)\nabla_{\rho}h\right\}\right]
\nonumber
\\
&\hskip 1 cm -2\left(\nabla_{\alpha}h^{\mu}_{\rho}-\frac{1}{2}\nabla^{\mu}h_{\alpha \rho}\right)\nabla^{(\alpha}\delta \xi^{\rho)}
\nonumber
\\
&\hskip 1 cm +\left(\nabla_{\alpha}\delta \mu\right)\left\{g^{\mu \alpha}\nabla_{\alpha}\left(\nabla_{\sigma}\delta \xi^{\sigma}\right)+\frac{1}{2}g^{\mu \alpha}\left[\left(\xi^{\rho}+\delta \xi^{\rho}\right)\nabla_{\rho}h\right]-\left(\nabla_{\sigma}\delta \xi^{\sigma}\right)h^{\mu \alpha}\right\}~.
\end{align}
%===============================================
As mentioned before, it will be useful if we write down simplified versions of Eq.~\ref{pake_form_04} for various scenarios of physical interest. These can range from the use of the transverse-traceless gauge to setting $\delta \mu=\textrm{constant}$. We discuss below each of these limits explicitly.
%===============================================
%===============================================
%===============================================
\begin{itemize}

\item[(i)] If we choose $\delta \mu=\textrm{constant}$, then the last term in the expression for $j^{\mu}$ in Eq.~\ref{pake_form_04} identically vanishes. In this context, the dynamical equation for the perturbed GKV becomes identical to that derived from the perturbation of the action, i.e., to Eq.~\ref{pake_action}.

\item[(ii)] If we assume that the background spacetime is vacuum, and the perturbation involves no incoming matter fields, then the use of the transverse-traceless gauge [equivalently setting $h=0$ in Eq.~\ref{pake_form_04}] yields
%===============================================
\begin{align}\label{reduction_ake_02}
\big(g^{\alpha \beta}&-h^{\alpha \beta}\big)\left[\nabla_{\alpha}\nabla_{\beta}\delta \xi^{\mu}+R^{\mu}_{~\alpha \rho \beta}\delta \xi^{\rho}\right]=j^{\mu}~,
\\
j^{\mu}&=-\left(1-\mu-\delta \mu\right)\left[\left(g^{\mu \alpha}-h^{\mu \alpha}\right)\nabla_{\alpha}\left(\nabla_{\sigma}\delta \xi^{\sigma}\right)
\right]-2\left(\nabla_{\alpha}h^{\mu}_{\rho}-\frac{1}{2}\nabla^{\mu}h_{\alpha \rho}\right)\nabla^{(\alpha}\delta \xi^{\rho)}
\nonumber
\\
&\hskip 1 cm +\left(\nabla_{\alpha}\delta \mu\right)\left\{g^{\mu \alpha}\nabla_{\alpha}\left(\nabla_{\sigma}\delta \xi^{\sigma}\right)-\left(\nabla_{\sigma}\delta \xi^{\sigma}\right)h^{\mu \alpha}\right\}~.
\end{align}
%===============================================
Note that any term involving $R_{\alpha \beta}$ will not contribute, since for vacuum spacetime the Ricci tensor identically vanishes. Also, if we assume $\delta \mu$ to be constant, the above equation can be simplified even more, as the last term in $j^{\mu}$ will be absent.

\item[(iii)] If we use the fact that the perturbed GKV is really a consequence of the perturbation of the spacetime geometry (we will examine this case in detail later), then we will have $\chi^{\mu}=\xi^{\mu}+\delta \chi^{\mu}_{1}+\delta\chi^{\mu}_{2}$, where $\delta \chi^{\mu}_{1}$ is linear in the gravitational perturbation, while $\delta \chi^{\mu}_{2}$ is quadratic in the gravitational perturbation. An identical decomposition will work for the $\bar{\mu}$ as well. If we keep terms linear in the gravitational perturbation, we should also ignore terms $\mathcal{O}(\delta \chi^{\mu}_{1}h_{\alpha \beta})$ and so on. It follows that perturbed AKE governing the evolution of the vector $\delta \chi^{\mu}_{1}$ takes the relatively simple form
%===============================================
\begin{align}\label{pake_formN_04}
\square \delta \chi^{\mu}_{1}+R{^\mu}_{\rho}\delta \chi^{\rho}_{1}+\left(1-\mu\right)\left[\nabla^{\mu}\left(\nabla_{\sigma}\delta \chi^{\sigma}_{1}\right)
+\frac{1}{2}\nabla^{\mu}\left(\xi^{\rho}\nabla_{\rho}h\right)\right]=0~.
\end{align}
%===============================================
Note also that if we assume the background spacetime to be vacuum with no incoming matter perturbation, then the use of the transverse-traceless gauge would reduce Eq.~\ref{pake_formN_04} to AKE for the background spacetime $g_{\mu \nu}$. The consequences of this equation with or without matter field will be discussed in a subsequent section.

\end{itemize}
%===============================================
%===============================================
%===============================================
Thus, we have derived the evolution equation for the perturbed GKV field from the perturbed action and also from the perturbed AKE to first order in the metric perturbations. Both of these procedures yield identical equations depicting the internal consistency of our analysis. We have also verified using computer algebra (in particular the xAct package for \textit{Mathematica}) that this consistency holds to second order as well. The AKE is rather complicated in the second-order case, so we do not present the result here; the interested reader can view the \textit{Mathematica} file posted at \cite{MathematicaRef}. In what follows, we will discuss the structure of the Hamiltonian associated with the dynamical equation for the perturbed AKE, leading to an understanding of the stability as well as its hyperbolicity.

%=======================================================================
%-----------------------------------------------------------------------
%
%		HYPERBOLICITY, HAMILTONIANS, AND STABILITY
%
%-----------------------------------------------------------------------
%=======================================================================
\section{Hamiltonian stability and Hyperbolicity}\label{pake_hamiltonian}

In this section, we will construct the Hamiltonian out of the Lagrangian, whose variation yields the evolution equation for the perturbed GKV. The stability of the Hamiltonian and its bounded nature will also be examined. In addition, the hyperbolicity of the perturbed AKE will also be explored.

%=======================================================
%-----------------------------------------------------------------------
%
%		Hamiltonian
%
%-----------------------------------------------------------------------
%=======================================================
\subsection{Hamiltonian for the perturbed AKE}

We have derived the evolution equation for the perturbed GKVs in the preceding section in two different ways: first by varying the perturbed action from which the AKE can be derived, and then by direct perturbation of the AKE. In this section, we will discuss the stability and the hyperbolicity of the perturbed AKE, restricting (for simplicity) to first order in the metric perturbations $h_{\mu \nu}$. First, we will construct the Hamiltonian for the perturbed GKV field and discuss its stability. Subsequently, we will discuss the hyperbolicity of the perturbed AKE and its consequences. The starting point for the Hamiltonian analysis is the action for the perturbed GKV field $\delta \xi^{\mu}$; in particular, the zeroth component of the boundary term in the variation will provide the Hamiltonian. The action for the perturbed GKV simplifies considerably to first order in the metric perturbations. Neglecting all terms of $\mathcal{O}(h_{\cdot \cdot}^{2})$ and using symmetry properties of the resulting expression, the structure of the action can be simplified to
%-unc (see \ref{AppD} for the derivation)
%===============================================
\begin{align}\label{perturbed_actionn}
\mathcal{A}[\delta \xi^\cdot]&=\int_{\mathcal{V}} d^{4}x\sqrt{-g}~\Big[\frac{1}{2}\left(1+\frac{1}{2}h\right)\left\{\bar{\mu}\delta^{\mu}_{\alpha}\delta ^{\nu}_{\beta}-g_{\alpha \beta}g^{\mu \nu}-h_{\alpha \beta}g^{\mu \nu}+g_{\alpha \beta}h^{\mu \nu}-\delta^{\nu}_{\alpha}\delta ^{\mu}_{\beta}\right\}\left(\nabla_{\mu}\delta \xi^{\alpha}\nabla_{\nu}\delta \xi^{\beta}\right)
\nonumber
\\
&+\frac{\bar{\mu}}{2}\Big\{\left(\xi^{\alpha}+\delta \xi^{\alpha}\right)\nabla_{\alpha}h\left(\nabla_{\mu}\delta \xi^{\mu}\right)\Big\}
-\left(\nabla_{\mu}\delta \xi^{\alpha}\right)\Big\{h_{\alpha \beta}\nabla^{\mu}\xi^{\beta}+h^{\mu \beta}\nabla_{\alpha}\xi_{\beta}\Big\}-g_{\nu \alpha}\nabla_{\mu}\delta \xi^{\alpha}\left(\xi^{\sigma}+\delta \xi^{\sigma}\right)\nabla_{\sigma}h^{\mu \nu}\Big]~.
\end{align}
%===============================================
Note that the Lagrangian density associated with the above action is identical to Eq.~\ref{lag_action_first}, though written in a different form.
%-unc The variation of the above action has been performed in detail in \ref{AppB};
Collecting all the total derivative terms that we have thrown away in the derivation of the field equation for $\delta \xi^{\mu}$ in the previous section, we obtain (recalling the notation $\Delta$ for the first-order variation)
%===============================================
\begin{align}
\textrm{Total~derivative~terms}&=\sqrt{-g}~\Delta \left(\delta \xi^{\alpha}\right)\Big[\left(\mu~ \delta^{\mu}_{\alpha}\delta ^{\nu}_{\beta}-g_{\alpha \beta}g^{\mu \nu}-\delta^{\nu}_{\alpha}\delta ^{\mu}_{\beta}\right)\nabla_{\nu}\delta \xi^{\beta}
\nonumber
\\
&\hskip -3 cm +\frac{1}{2}~h\left(\mu~ \delta^{\mu}_{\alpha}\delta ^{\nu}_{\beta}-g_{\alpha \beta}g^{\mu \nu}-\delta^{\nu}_{\alpha}\delta ^{\mu}_{\beta}\right)\nabla_{\nu}\delta \xi^{\beta}
+\left(1+\frac{1}{2}h\right)\delta \mu~ \delta^{\mu}_{\alpha}\delta ^{\nu}_{\beta}\nabla_{\nu}\delta \xi^{\beta}
+\left(g_{\alpha \beta}h^{\mu \nu}-h_{\alpha \beta}g^{\mu \nu}\right)\nabla_{\nu}\delta \xi^{\beta}
\nonumber
\\
&\hskip -3 cm +\frac{\bar{\mu}}{2}\Big\{\left(\xi^{\sigma}+\delta \xi^{\sigma}\right)\nabla_{\sigma}h\Big\}\delta^{\mu}_{\alpha}
-\Big\{h_{\alpha \beta}\nabla^{\mu}\xi^{\beta}+h^{\mu \beta}\nabla_{\alpha}\xi_{\beta}\Big\}
-g_{\nu \alpha}\left(\xi^{\sigma}+\delta \xi^{\sigma}\right)\nabla_{\sigma}h^{\mu \nu}\Big]
\equiv \sqrt{-g}~\Delta \left(\delta \xi^{\alpha}\right)~P{^\mu}_{\alpha}~,
\end{align}
%===============================================
where the last equality defines the quantity $P{^\mu}_{\alpha}$, which is the polymomentum conjugate to the perturbed GKV field $\delta \xi^{\alpha}$.
In particular, starting from the perturbed action presented in Eq.~\ref{perturbed_actionn}, one can immediately verify that $P{^\mu}_{\alpha}$ has the following expression in terms of the perturbed Killing vector field $\delta \xi^{\alpha}$ and its derivatives:
%===============================================
\begin{align}\label{momentum}
P{^\mu}_{\alpha}&\equiv \frac{1}{\sqrt{-g}}\frac{\Delta \mathcal{A}}{\Delta(\nabla_{\mu}\delta \xi^{\alpha})}
\nonumber
\\
&=\Big[\left(1+\frac{1}{2}h\right)\left(\bar{\mu}~ \delta^{\mu}_{\alpha}\delta ^{\nu}_{\beta}-g_{\alpha \beta}g^{\mu \nu}-\delta^{\nu}_{\alpha}\delta ^{\mu}_{\beta}+g_{\alpha \beta}h^{\mu \nu}-h_{\alpha \beta}g^{\mu \nu}\right)\nabla_{\nu}\delta \xi^{\beta}
\nonumber
\\
&\hskip 2 cm +\frac{\bar{\mu}}{2}\Big\{\left(\xi^{\sigma}+\delta \xi^{\sigma}\right)\nabla_{\sigma}h\Big\}\delta^{\mu}_{\alpha}
-\Big\{h_{\alpha \beta}\nabla^{\mu}\xi^{\beta}+h^{\mu \beta}\nabla_{\alpha}\xi_{\beta}\Big\}
-g_{\nu \alpha}\left(\xi^{\sigma}+\delta \xi^{\sigma}\right)\nabla_{\sigma}h^{\mu \nu}\Big]~.
\end{align}
%===============================================
We write the variation of the action for the perturbed Killing vector field $\delta \xi^{\alpha}$ (incorporating the variation of the boundary surface) in the Weiss form \cite{FengMatznerWeiss2018,*SudarshanCM,*MatznerShepleyCM,*Weiss1936}:
%===============================================
\begin{align}
\Delta \mathcal{A}=\int _{\mathcal{V}}d^{4}x\sqrt{-g}~E_{\alpha}~\Delta \left(\delta \xi^{\alpha}\right)+\int _{\partial \mathcal{V}}d\Sigma_{\mu}\left[P{^\mu}_{\alpha}~\Delta \left(\delta \xi^{\alpha}\right)+L\Delta x^{\mu}\right]~,
\end{align}
%===============================================
where $d\Sigma_{\alpha}=d^{3}x\sqrt{-g}\nabla_{\alpha}\phi$ is the volume measure on the boundary surface $\partial \mathcal{V}$ denoting the $\phi(x^{\mu})=\textrm{constant}$ hypersurface. If we instead use the unit normal vector $n_{\alpha}$, then the volume measure of the boundary hypersurface $\partial \mathcal{V}$ becomes $d\Sigma_{\alpha}=\epsilon d^{3}x\sqrt{h}n_{\alpha}$, where $\epsilon=-1(+1)$ for spacelike (timelike) hypersurfaces, respectively. Defining $d\Sigma=d^{3}x\sqrt{h}$ and choosing the hypersurface to be $t=\textrm{constant}$ and using the $(1+3)$ decomposition for the metric, we find the boundary Hamiltonian to be
%===============================================
\begin{align}
H=\int d\Sigma~ \mathcal{H}~,\qquad \mathcal{H}=\epsilon P{^\mu}_{\alpha}n_{\mu}\dot{\delta \xi}^{\alpha}-NL~,
\end{align}
%===============================================
where, $N$ is the lapse function and the ``overdot'' denotes derivative with respect to time. For the spacelike hypersurface we are considering, and using orthonormal coordinates, e.g., in the synchronous frame, the Hamiltonian density $\mathcal{H}$ takes the following form:
%===============================================
\begin{align}
\mathcal{H}=P{^0}_{\alpha}\dot{\delta \xi}^{\alpha}-\mathcal{L}~.
\end{align}
%===============================================
We now consider terms which are quadratic in the time derivative as well as terms which are quadratic in the spatial derivative of the perturbed GKV field $\delta \xi^{\alpha}$. Collecting these terms, the time derivative part of the Hamiltonian density quadratic in the perturbed Killing vector field becomes
%===============================================
\begin{align}\label{Hamiltonian_time}
\mathcal{H}^{\rm (2) time}=\frac{1}{2}\left(1+\frac{1}{2}h\right)\left\{\bar{\mu}-2\right\}\left(\dot{\delta \xi}^{0}\dot{\delta \xi}^{0}\right)
+\frac{1}{2}\left(1+\frac{1}{2}h\right)\left\{g_{ij}+h_{ij}+g_{ij}h^{00}\right\}\left(\dot{\delta \xi}^{i}\dot{\delta \xi}^{j}\right)
+\left(1+\frac{1}{2}h\right)h_{0i}\left(\dot{\delta \xi}^{0}\dot{\delta \xi}^{i}\right)~.
\end{align}
%===============================================
Here we have performed a (1+3) decomposition of the Hamiltonian density and have collected terms quadratic in the time derivative of the perturbed GKV field. In the limit of vanishing perturbation, the above quadratic contribution to the Hamiltonian coincides with that presented in \cite{FGW2019}. Though the kinetic term of the zeroth component of the perturbed Killing vector field, i.e., $\delta \xi^{0}$ in the Hamiltonian density harbors a negative sign in the unperturbed spacetime for $\mu<2$ (see \cite{FGW2019}), we see that in Eq.~\ref{Hamiltonian_time}, the corresponding kinetic term in perturbed GKV has positive sign for $\mu<2<\bar{\mu}$. Thus ghost modes can be avoided. The other quadratic terms in the time derivative have positive sign. Similarly, terms quadratic in the space derivatives of the perturbed Killing vector field yield the following expression for the Hamiltonian density:
%===============================================
\begin{align}\label{Hamiltonian_space}
\mathcal{H}^{\rm (2) space}&=\frac{1}{2}\left(1+\frac{1}{2}h\right)\left\{1-\bar{\mu}\right\}\left(\partial_{i}\delta \xi^{i}\partial_{j}\delta \xi^{j}\right)
+\frac{1}{2}\left(1+\frac{1}{2}h\right)\left\{-g^{ij}+h_{00}g^{ij}+h^{ij}\right\}\left(\partial_{i}\delta \xi^{0}\partial_{j}\delta \xi^{0}\right)
\nonumber
\\
&+\frac{1}{2}\left(1+\frac{1}{2}h\right)\left\{g_{ab}g^{ij}+h_{ab}g^{ij}-g_{ab}h^{ij}\right\}\left(\partial_{i}\delta \xi^{a}\partial_{j}\delta \xi^{b}\right)+\left(1+\frac{1}{2}h\right)h_{0a}g^{ij}\left(\partial_{i}\delta \xi^{0}\partial_{j}\delta \xi^{a}\right)~.
\end{align}
%===============================================
In the above expression involving spatial derivatives of the perturbed GKV, the first two terms can provide a negative contribution to the Hamiltonian, thereby making it unbounded from below. In the second term, even though the metric perturbations try to keep this term positive, the background metric drives it to negative values. Similarly, if we want the theory to be ghost-free, the first term will turn negative, leading to an unbounded Hamiltonian. As argued in \cite{FGW2019}, in the unperturbed case, an unbounded Hamiltonian is potentially dangerous, as it can result in runaway behavior that drives the GKVs far from the Killing condition. Thus, the problems associated with the Hamiltonian density for the AKE in the exact case remain for the perturbed GKV $\delta \xi^{\alpha}$ as well.
%=======================================================
%-----------------------------------------------------------------------
%
%		Hyperbolicity
%
%-----------------------------------------------------------------------
%=======================================================
\subsection{Hyperbolicity}

We now turn our attention to the hyperbolicity of the perturbed AKE. For this purpose, we employ the methods of hyperbolicity analysis for second-order systems, particularly that presented in \cite{Gun06,Hil13} (see also \cite{FGW2019}). In this approach, we compute the principal symbol for the system of equations; if the principal symbol has real eigenvalues, the system is weakly hyperbolic, and if the principal symbol has a complete set of eigenvectors, the system is strongly hyperbolic. Collecting all the terms involving double derivatives of $\delta \xi^{\alpha}$, we obtain from Eq.~\ref{pake_form_04},
%===============================================
\begin{align}\label{pake_hyp_01}
\left(g^{\alpha \beta}-h^{\alpha \beta}\right)\partial_{\alpha}\partial_{\beta}\delta \xi^{\mu}+\left(1-\mu-\delta \mu\right)\left(g^{\mu \alpha}-h^{\mu \alpha}\right)\partial_{\alpha}\partial_{\sigma}\delta \xi^{\sigma}\approx 0~,
\end{align}
%===============================================
where the symbol $\approx$ denotes equality up to terms not included in the principal part. In order to express the above equation in the desired form, we can decompose the metric as $g^{\alpha \beta}=q^{\alpha \beta}-n^{\alpha}n^{\beta}+s^{\alpha}s^{\beta}$, where $n^{\alpha}$ is a timelike unit vector and $s^{\alpha}$ is a spacelike unit vector. Further, defining $n^{\alpha}\partial_{\alpha}=\partial_{n}$, $s^{\alpha}\partial_{\alpha}=\partial_{s}$, and $q^{\alpha \beta}=q^{AB}\delta^{\alpha}_{A}\delta^{\beta}_{B}$, we can rewrite the above equation into three separate equations; we obtain one by contraction of Eq.~\ref{pake_hyp_01} with $n_{\mu}$, another by contraction of Eq.~\ref{pake_hyp_01} with $s_{\mu}$, and the last by the projection of Eq.~\ref{pake_hyp_01} along transverse directions. Keeping only the principal parts of these equations, we obtain
%-unc (for the derivation, see \ref{AppE})
%===============================================
\begin{align}\label{hypb_04}
\left(2-\mu-\delta \mu\right)\left(1+h^{nn}\right)\partial_{n}^{2}\delta \xi^{n}&\approx\left(1-h^{ss}\right)\partial_{s}^{2}\delta \xi^{n}
+\left(3-\mu-\delta \mu\right)h^{ns}\partial_{n}\partial_{s}\delta \xi^{n}
\nonumber
\\
&\hskip1 cm +\left(1-\mu-\delta \mu\right)\left(1+h^{nn}\right)\partial_{n}\partial_{s}\delta \xi^{s}-\left(1-\mu-\delta \mu\right)h^{n s}\partial_{s}^{2}\delta \xi^{s}~.
\end{align}
%===============================================
Note that since $n^{\mu}$ is the timelike unit vector, $\partial_{n}^{2}\delta \xi^{n}$ corresponds to a double time derivative of the time component of the perturbed GKV. A similar analysis yields the following equation for $\partial_{n}^{2}\delta \xi^{s}$, i.e., for the double time derivative of the spatial component of the perturbed GKV,
%===============================================
\begin{align}\label{hypb_06}
\left(1+h^{nn}\right)\partial_{n}^{2}\delta \xi^{s}&\approx\left(2-\mu-\delta \mu\right)\left(1-h^{ss}\right)\partial_{s}^{2}\delta \xi^{s}
+\left(3-\mu-\delta \mu\right)h^{ns}\partial_{n}\partial_{s}\delta \xi^{s}
\nonumber
\\
&\hskip 1 cm -\left(1-\mu-\delta \mu\right)h^{sn}\partial_{n}^{2}\delta \xi^{n}
-\left(1-\mu-\delta \mu\right)\left(1-h^{ss}\right)\partial_{s}\partial_{n}\delta \xi^{n}~.
\end{align}
%===============================================
Finally, the double time derivative of the transverse component of the perturbed GKV yields
%===============================================
\begin{align}\label{hypb_08}
\left(1+h^{nn}\right)\partial_{n}^{2}\delta \xi^{A}&\approx \left(1-h^{ss}\right)\partial_{s}^{2}\delta \xi^{A}
+2h^{ns}\partial_{n}\partial_{s}\delta \xi^{A}
+\left(1-\mu-\delta \mu\right)h^{A s}\partial_{s}\partial_{n}\delta \xi^{n}
\nonumber
\\
&\hskip -1 cm -\left(1-\mu-\delta \mu\right)h^{A n}\partial_{n}^{2}\delta \xi^{n}
+\left(1-\mu-\delta \mu\right)h^{A n}\partial_{n}\partial_{s}\delta \xi^{s}
-\left(1-\mu-\delta \mu\right)h^{A s}\partial_{s}^{2}\delta \xi^{s}~.
\end{align}
%===============================================
Therefore, we can read off the principal symbol $P^{s}$ for this system of second-order differential equations, which takes the following form:
%-unc (again see \ref{AppE} for the derivation)
%===============================================
\begin{align}
P^{s}=
\begin{bmatrix}
O_{4\times 4} & I_{4\times 4}\\
\mathbb{A} 	              & \mathbb{B}
\end{bmatrix}~,
\end{align}
%===============================================
where $O_{4\times 4}$ and $I_{4\times 4}$ are the $(4\times 4)$ null and unit matrix respectively. The entries $\mathbb{A}$ and $\mathbb{B}$ are also $(4\times 4)$ matrices with the following expressions:
%===============================================
\begin{align}
\mathbb{A}=
\begin{bmatrix}
\mathcal{A} & \mathcal{B} & 0                 \\
\mathcal{E} & \mathcal{F} & 0                 \\
\mathcal{I}^{A}  & \mathcal{J}^{A}  & \mathcal{K}I_{2\times 2}\\
\end{bmatrix}~,
\quad
\mathbb{B}=
\begin{bmatrix}
\mathcal{C} & \mathcal{D} & 0 \\
\mathcal{G} & \mathcal{H} & 0 \\
\mathcal{L}^{A} & \mathcal{M}^{A} & \mathcal{N}I_{2\times 2} \\
\end{bmatrix}~,
\end{align}
%===============================================
where the unknown quantities appearing in the matrices $\mathbb{A}$ and $\mathbb{B}$ have the following expressions,
%===============================================
\begin{align}
\mathcal{A}&=\frac{\left(1-h^{ss}\right)}{\left(2-\mu-\delta \mu\right)\left(1+h^{nn}\right)}~,
\quad
\mathcal{B}=-\frac{\left(1-\mu-\delta \mu\right)h^{ns}}{\left(2-\mu-\delta \mu\right)\left(1+h^{nn}\right)}~,
\\
\mathcal{C}&=\frac{\left(3-\mu-\delta \mu\right)h^{ns}}{\left(2-\mu-\delta \mu\right)\left(1+h^{nn}\right)}~,
\quad
\mathcal{D}=\frac{\left(1-\mu-\delta \mu\right)}{\left(2-\mu-\delta \mu\right)}~,
\\
\mathcal{E}&=-\frac{\left(1-\mu-\delta \mu\right)h^{sn}}{\left(2-\mu-\delta \mu\right)\left(1+h^{nn}\right)^{2}}~,
\quad
\mathcal{F}=\frac{\left(2-\mu-\delta \mu\right)\left(1-h^{ss}\right)}{\left(1+h^{nn}\right)}~,
\\
\mathcal{G}&=-\frac{\left(1-\mu-\delta \mu\right)\left(1-h^{ss}\right)}{\left(1+h^{nn}\right)}~,
\quad
\mathcal{H}=\frac{\left(3-\mu-\delta \mu\right)h^{ns}}{\left(1+h^{nn}\right)}
-\frac{\left(1-\mu-\delta \mu\right)^{2}h^{sn}}{\left(2-\mu-\delta \mu\right)\left(1+h^{nn}\right)}~,
\\
\mathcal{I}&=-\frac{\left(1-\mu-\delta \mu\right)h^{An}}{\left(2-\mu-\delta \mu\right)\left(1+h^{nn}\right)^{2}}~,
\quad
\mathcal{J}=-\frac{\left(1-\mu-\delta \mu\right)h^{As}}{\left(1+h^{nn}\right)}~,
\\
\mathcal{K}&=\frac{\left(1-h^{ss}\right)}{\left(1+h^{nn}\right)}~,
\quad
\mathcal{L}=\frac{\left(1-\mu-\delta \mu\right)h^{As}}{\left(1+h^{nn}\right)}~,
\\
\mathcal{M}&=-\frac{\left(1-\mu-\delta \mu\right)^{2}h^{An}}{\left(2-\mu-\delta \mu\right)\left(1+h^{nn}\right)}
+\frac{\left(1-\mu-\delta \mu\right)h^{An}}{\left(1+h^{nn}\right)}~,
\quad
\mathcal{N}=2\frac{h^{ns}}{\left(1+h^{nn}\right)}.
\end{align}
%===============================================
Note that as the perturbations vanish, we obtain $\mathcal{A}=1/(2-\mu)$, $\mathcal{B}=0=\mathcal{C}$, $\mathcal{D}=(1-\mu)/(2-\mu)$, $\mathcal{E}=0$, $\mathcal{F}=2-\mu$, $\mathcal{G}=-(1-\mu)$, $\mathcal{H}=0=\mathcal{I}=\mathcal{J}$, $\mathcal{K}=1$, and $\mathcal{L}=0=\mathcal{M}=\mathcal{N}$. Thus, our result agrees with that derived in \cite{FGW2019}. If one chooses $\mu + \delta \mu = 1$, one has a complete set of eigenvectors, and if one chooses $\mu + \delta \mu = 2$, the principal symbol becomes singular. Setting $\mu=1$ and expanding the eigenvalues to first order in metric perturbations and second order in $\delta \mu$, one obtains the following expression for the eigenvalues of the principal symbol:\footnote{These calculations were performed using the package xAct in \textit{Mathematica}.}
%===============================================
\begin{equation}
   \begin{aligned}
      \biggl\{
      &
      \frac{1}{2} (h^{nn}-2 h^{ns}+h^{ss}-2)
      ,
      -\frac{1}{2} (h^{nn}+2 h^{ns}+h^{ss}-2)
      ,
      \frac{1}{4} \left(-\delta \mu^2 h^{ns}+\left(\delta \mu^2+4\right) h^{ns}-2 h^{nn}-2 (h^{ss}-2)\right)
      , \\
      &
      \frac{1}{4} \left(-\delta \mu^2 h^{ns}+\left(\delta \mu^2+4\right) h^{ns}-2 h^{nn}-2 (h^{ss}-2)\right)
      ,
      \frac{1}{4} \left(-\delta \mu^2 h^{ns}+\left(\delta \mu^2+4\right) h^{ns}+2 h^{nn}+2 (h^{ss}-2)\right)
      , \\
      &
      \frac{1}{4} \left(-\delta \mu^2 h^{ns}+\left(\delta \mu^2+4\right) h^{ns}+2 h^{nn}+2 (h^{ss}-2)\right)
      \biggr\}
   \end{aligned}
\end{equation}
%===============================================
Since the eigenvalues presented above are real, this implies that to first order in the metric perturbation and second order in $\delta \mu$, the perturbed AKE is weakly hyperbolic. This is consistent with the result in \cite{FGW2019}, in which the AKE for the background spacetime was found to be weakly hyperbolic for general $\mu \neq 1,2$.

%=======================================================================
%-----------------------------------------------------------------------
%
%		INDUCED PERTURBATIONS
%
%-----------------------------------------------------------------------
%=======================================================================
\section{Induced perturbations: Implications for the almost Killing equation}\label{pake_sol}

So far, we had considered the metric perturbation and the perturbation of the GKV to be independent. However, in most of the physical scenarios of interest, e.g., perturbation of the black hole spacetime due to matter fields entering the horizon, the metric perturbation sources the perturbation of the almost Killing equation. Thus, we may consider the perturbation of the GKV to be induced by the metric perturbation. In what follows, we will consider such induced perturbation of the GKV and hence determine their evolution equations order by order.
%===============================================
%-----------------------------------------------------------------------
%
%		First Order
%
%-----------------------------------------------------------------------
%===============================================
\subsection{First-order perturbation}\label{strict_first}

As emphasized before, we will imagine a class of nontrivial perturbative solutions to the AKE which are induced by the metric perturbations. This may be quantified by assuming that $\delta \xi^\alpha$ and $h_{\alpha \beta}$ are implicitly proportional to the same expansion parameter $\epsilon$ (see footnote 2), and then solving the perturbed AKE order by order in $\epsilon$. In particular, for the GKV $\chi^\alpha$ we write
%===============================================
\begin{equation} \label{GKVPert}
\chi^{\alpha} = \xi^{\alpha} + \delta \chi^{\alpha}_1 + \delta \chi^{\alpha}_2 + \mathcal{O}(\epsilon^3)~,
\end{equation}
%===============================================
where we have assumed $\delta \chi^{\alpha}_i \propto \epsilon^i$. As before, the background spacetime is assumed to admit a Killing vector field $\xi^{\alpha}$, which satisfies the AKE presented in Eq.~\ref{AlmostKillingEquation} exactly. Further, imposing the Lorenz gauge condition, $\nabla^\beta h_{\alpha \beta} = \nabla_\alpha h / 2$, and choosing the background value of the parameter $\mu$ to be constant, the expansion of the perturbed AKE presented in Eq.~\ref{pake_form_04} to $\mathcal{O}(\epsilon)$ reduces to the following equation for $\delta \chi^{\alpha}_1$:
%===============================================
\begin{equation} \label{AKEPertI}
\Box \delta \chi^{\alpha}_1
+ R{^\alpha}{_\beta} \, \delta \chi^{\beta}_1
+ (1 - \mu)\nabla^\alpha \left[\nabla \cdot \delta \chi_1 + \left(\xi^\beta \, \nabla_\beta h \right) / 2 \right]
= 0~.
\end{equation}
%===============================================
Note that this reduces to the background AKE when $\mu=1$ or, when $\xi^\beta \, \nabla_\beta h$ is a constant; in those cases, one must consider higher-order corrections to the AKE. In the cases where $\xi^\beta \, \nabla_\beta h$ is nontrivial, for instance, if the perturbations are the result of an energy-momentum tensor with nontrivial trace, one can apply the expression Eq.~\ref{NoetherCurrent} for the Noether current defined with respect to the background derivatives directly to $\delta \chi^{\alpha}_1$ and substitute Eq.~\ref{AKEPertI} to obtain the following current:
%===============================================
\begin{equation} \label{KomarPAKE1}
\begin{aligned}
J^\alpha_1
    %    &= \nabla_\nu \left( \nabla^\mu \delta \chi^\nu_1 - \nabla^\nu \delta \chi^\mu_1 \right) \\
    %    &=  \nabla_\nu \nabla^\mu \delta \chi^\nu_1 - \Box \delta \chi^\mu_1 \\
    %    &=  \nabla_\nu \nabla^\mu \delta \chi^\nu_1 + R{^\mu}{_\sigma} \, \delta \chi^{\sigma}_1 + \nu \nabla^\mu \left(\nabla \cdot \delta \chi_1 + \xi^\alpha \nabla_\alpha h / 2 \right) \\
    %    &=  \nabla_\nu \nabla^\mu \delta \chi^\nu_1 - \nabla^\mu \nabla_\nu \delta \chi^\nu_1 + \nabla^\mu (\nabla \cdot \delta \chi_1) + R{^\mu}{_\sigma} \, \delta \chi^{\sigma}_1
    %    + \nu \nabla^\mu \left(\nabla \cdot \delta \chi_1 + \xi^\alpha \nabla_\alpha h / 2 \right) \\
&= 2 R{^\alpha}{_\beta} \, \delta \chi^{\beta}_1
+ \nabla^\alpha \left[\left(2-\mu\right)\nabla \cdot \delta \chi_1 + \left(1-\mu\right)\, \xi^\alpha \nabla_\alpha h / 2 \right]~,
\end{aligned}
\end{equation}
%===============================================
which satisfies the identity $\nabla \cdot J_1=0$. If the background spacetime is vacuum, setting $R{^\alpha}_{\beta}=0$, this identity yields
%===============================================
\begin{equation} \label{KomarPAKEPropagation}
\Box \Phi_1 = 0~,
\end{equation}
%===============================================
where
%===============================================
\begin{equation} \label{Constraint}
\Phi_1 := \left(2-\mu\right)\nabla \cdot \delta \chi_1 +\frac{1}{2} \left(1-\mu\right)\, \left(\xi^\alpha \nabla_\alpha h\right)~.
\end{equation}
%===============================================
It follows that if the initial data satisfy the constraint $\Phi_1=0$ and its first time derivative also vanishes, then the constraint $\Phi_{1}=0$ is preserved by the evolution of Eq.~\ref{KomarPAKEPropagation}. Under this constraint, Eq.~\ref{AKEPertI} may be rewritten as
%===============================================
\begin{equation} \label{AKEPertIB}
\begin{aligned}
		& \Box \delta \chi^{\alpha}_1 - \nabla^\alpha \left(\nabla \cdot \delta \chi_1 \right) = 0~ ,
\end{aligned}
\end{equation}
%===============================================
which is equivalent to the AKE for $\mu = 2$ in vacuum spacetime. It was demonstrated in \cite{FGW2019} that the $\mu = 2$ parameter choice avoids ghosts and is dynamically equivalent to the Maxwell theory, thus avoiding the problems arising from an unbounded Hamiltonian.

Furthermore, recall that to first order in the gravitational perturbation, imposing the Lorenz gauge condition, the trace $h$ of the gravitational perturbation satisfies the following evolution equation on a vacuum background:
%===============================================
\begin{equation} \label{EFEPERTtr}
\Box h = 16 \pi \delta_1 T~,
\end{equation}
%===============================================
where $\delta_{1}T$ is the first-order perturbation of the trace of the matter energy-momentum tensor. If one assumes $\delta_1 T = 0$, i.e., the matter field perturbing the spacetime is dilute radiation and the background is vacuum, one can impose the condition that $h = 0$, in which case, from Eq.~\ref{Constraint}, it follows that the constraint $\Phi_1=0$ implies $\nabla \cdot \delta \chi_1=0$. Subsequently, substituting this result in Eq.~\ref{AKEPertIB} simplifies to $\Box \delta \chi^\alpha_1=0$, which is satisfied by the background Killing vector field $\xi^{\alpha}$. Thus, one may expect that in this case, the perturbed spacetime will respect the symmetries of the background spacetime. On the other hand, if $\delta_{1}T \neq 0$, then the background Killing vector field does not, in general, satisfy the constraint $\Phi_1=0$, and the resulting solutions for Eq.~\ref{AKEPertIB} differ nontrivially from the background Killing vector.

We have therefore derived that if the spacetime perturbation $h_{\alpha \beta}$ is being sourced by an energy-momentum tensor with a nonvanishing trace, then the evolution equation for the first-order perturbation $\delta \chi^{\alpha}_1$ must differ from the background Killing vector field in a nontrivial manner. Furthermore, we find that for the weakly hyperbolic parameter choices, i.e., $\mu \neq 1$ and $\mu \neq 2$, and initial data satisfying the constraint $\Phi_1 = 0$ and $\partial_{t}\Phi_{1}=0$, the AKE propagates the constraint $\Phi_1 = 0$ and avoids the dynamical instabilities associated with ghosts and unboundedness in the Hamiltonian. First-order perturbations of the AKE are therefore different from the background Killing vector field and are well suited for describing the perturbations of the vacuum spacetimes induced by dilute matter in which $\delta_{1} T \neq 0$.

%=======================================================================
%-----------------------------------------------------------------------
%
%		SECOND-ORDER PERTURBATIONS
%
%-----------------------------------------------------------------------
%=======================================================================
\subsection{Second-order perturbations}

As the previous section demonstrates, perturbations of the AKE to $\mathcal{O}(\epsilon)$ are insensitive to the metric perturbations in scenarios with $\delta_{1}T=0$. This is the case if the perturbations consist of weak gravitational radiation, or dilute matter source encompassing radiation or null dust. In such situations, one should consider the evolution equation for the second-order correction $\delta \chi^{\nu}_2$. Assuming constant $\mu$ and a vacuum background spacetime, the homogeneous part of the equation is the same as that of the first-order case
%===============================================
\begin{equation} \label{AKEPertII}
\Box \delta \chi^{\nu}_2 + (1 - \mu) \nabla^\nu \left(\nabla \cdot \delta \chi_2 \right) = - \left( j_2^\nu + k_2^\nu + l_2^\nu + m_2^\nu \right)~,
\end{equation}
%===============================================
and the inhomogeneous part $j_2^\nu$ takes the explicit form
%===============================================
\begin{equation} \label{AKEPertIIj}
\begin{aligned}
j^\nu_2
= \,
&
    2 \xi{^\alpha} h^{\beta \sigma}
    \{
        h^{\nu \tau} R_{\alpha \beta \tau \sigma }
        -
        h{_\beta}{^\tau} R{^\nu}_{\sigma \alpha \tau}
        -
        \nabla{_\sigma} \nabla{_\alpha} h{^\nu}{_\beta}
    \}
    -
    \xi{^\alpha} \nabla{_\alpha} h_{\beta \sigma} \nabla{^\nu} h^{\beta \sigma}
    +
    2 \nabla{^\nu}
    (
        \xi{^\alpha} h^{\beta \sigma} \nabla{_\alpha} h_{\beta \sigma}
    )
    -
    2 h^{\beta \sigma} \nabla{^\nu} \xi{^\alpha} \nabla{_\sigma} h_{\alpha \beta}
\\
&
    +
    2 h{_\alpha}{^\sigma} \nabla{^\beta}\xi{^\alpha}
    \{
        \nabla{_\beta} h{^\nu}{_\sigma}
        -
        \nabla{^\nu} h_{\beta \sigma}
        +
        \nabla{_\sigma} h{^\nu}{_\beta}
    \}~,
\\
k^\nu_2
= \,
&
    2 h{^\nu}{_\beta} \nabla{^\beta} \nabla{_\alpha} \delta \chi_1^\alpha
    -
    2 h_{\alpha \beta} \nabla{^\beta} \nabla{^\alpha} \delta \chi_1^\nu
    +
    2 \nabla{^\beta} \delta \chi_1^\alpha
    \{
        \nabla{_\alpha} h{^\nu}{_\beta}
        +
        \nabla{_\beta} h{^\nu}{_\alpha}
        -
        \nabla{^\nu} h_{\alpha \beta}
    \}
    -
    2 \delta \chi_1^\alpha h^{\beta \sigma} R{^\nu}_{\beta \alpha \sigma}~,
\\
l^\nu_2
= \,
&
    h^{\nu \beta} \nabla{_\beta} \xi{^\alpha} \nabla{_\alpha}h
    -
    \nabla{^\nu}
    \left[
        \delta \chi_1^\alpha \nabla{_\alpha}h
        +
        \left(
            2 \nabla{_\alpha} \delta \chi_1^\alpha
            +
            \xi{^\alpha} \nabla{_\alpha}h
        \right)
        \delta \mu
    \right]
    +
    (\mu-2)
    \{
        h^{\nu \beta} \nabla{_\beta}
        (
            \xi{_\alpha} \nabla{^\alpha}h
        )
        -
        \nabla{^\nu}
        (
            \delta \chi_1^\alpha \nabla{_\alpha}h
        )
    \}~,
\\
m^\nu_2
= \,
&
    (\mu-2)
    \{
        \nabla{^\nu}
        (
            \xi{^\alpha} h^{\beta \sigma} \nabla{_\alpha} h_{\beta \sigma}
        )
        +
        2 h{^\nu}{_\beta} \nabla{^\beta} \nabla{_\alpha} \delta \chi_1^\alpha
    \}~.
\end{aligned}
\end{equation}
%===============================================
The terms in Eq.~\ref{AKEPertIIj} are organized so that $k_2^\nu = 0$, if $\delta \chi^\mu_1 = 0$ or, if $\delta \chi^\mu_1$ satisfies the Killing equation, and $l_2^\nu = 0$ if $h=0=\delta \mu$. As pointed out in the first-order case, for $h=0$, the relevant solutions for $\delta \chi_{1}^{\alpha}$ are essentially those of the background Killing vector; in that case, one may choose $\delta \chi^\mu_1=0$, as one can absorb it into the background Killing vector field.

Further understanding of the second-order perturbation can be achieved by computing the second-order perturbation of the Noether current assuming constant $\mu$ and a vacuum background $R_{\mu \nu} = 0$:
%===============================================
\begin{equation} \label{AKEPertII-Komar-full}
\begin{aligned}
J_2^\nu =
&	\,
    4 \delta \chi_1^\alpha \, \delta_1 R{^\nu}{_\alpha}
    +
    2 \xi^\alpha
    \left(
        \delta_2 R{^\nu}{_\alpha}
        -
        2h{^\nu}{_\beta} \delta_1 R{_\alpha}{^\beta}
    \right)
    +
    (\mu-2)
    \biggl[
        \nabla_\alpha h_{\sigma \tau}
        (
            \xi^\alpha \nabla^\nu h^{\sigma \tau}
            +
            h^{\sigma \tau} \nabla^\nu \xi^\alpha
        )
        +
        \xi^\alpha  h^{\sigma \tau} \nabla^\nu \nabla_\alpha h_{\sigma \tau}
\\
& \,
        +
        2 h{^\nu}{_\beta} \nabla^\beta \left(\nabla \cdot \delta \chi_1 \right)
        -
        \nabla^\nu \left(\nabla \cdot \delta \chi_2 \right)
    -
    \left\{
        \nabla^\nu \left(\delta \chi_1^\alpha \nabla_\alpha h\right)
        -
        h^{\nu \beta} \nabla_\beta \left(\xi^\alpha \nabla_\alpha h\right)
    \right\}
    \biggr]
	~,
\end{aligned}
\end{equation}
%===============================================
where $\delta_1 R_{\alpha \beta}$ and $\delta_2 \, R_{\alpha \beta}$ are the respective first- and second-order perturbations of the lowered index Ricci tensor. Note that the quantity within the curly brackets vanishes when $h=0$. We further see that when $\mu=2$, most of the terms in the Noether current, except for the first two terms, identically vanish. Therefore, for $\mu=2$, the Noether current depends on the perturbations of the Ricci tensor alone, as one might expect.

We then turn to the second-order perturbations of the identity presented in Eq.~\ref{KomarAKcond}. Assuming vacuum spacetime along with transverse-traceless gauge $h=0$, and setting $\mu$ to be a constant, the perturbation of the Komar identity takes the following form:
%===============================================
\begin{equation} \label{AKEPertII-KomarAKcond}
\begin{aligned}
\left(\mu-2\right) \Psi_{\rm L}=\Psi_{\rm R}~,
\end{aligned}
\end{equation}
%===============================================
where
%===============================================
\begin{equation} \label{AKEPertII-KomarAKcondL}
\begin{aligned}
\Psi_{\rm L}\equiv &
    ~\Box
    \left[
      \nabla \cdot \delta \chi_2
      +
      \delta \chi_1^\alpha \nabla_\alpha h
      -
      h^{\sigma \tau} \nabla_\alpha
      \left(
        \xi^\alpha h_{\sigma \tau}
      \right)
    \right]
    +
    2 h^{\sigma \tau}
    \left(
      \nabla_\nu\nabla_\sigma\nabla_\tau\delta \chi_1^\nu
      +
      R_{\alpha \sigma \tau \beta } \nabla^\beta\delta \chi_1^\alpha
    \right)\\
    &
    + \nabla_\beta \left[\nabla_\alpha (h^{\beta \alpha} \xi^\sigma \nabla_\sigma h) \right] + \nabla^\beta h \nabla_\beta\left(\xi^\alpha \nabla_\alpha h\right)~,
\end{aligned}
\end{equation}
%===============================================
%===============================================
\begin{equation} \label{AKEPertII-KomarAKcondR}
\begin{aligned}
\Psi_{\rm R}&\equiv
      ~\xi^\alpha
      \left(
          \nabla_\alpha \delta_2 R
          -
          2 h^{\sigma \tau} \nabla_\alpha \delta_1 R_{\sigma \tau}
      \right)
      -
      4 \delta_1 R_{\alpha \sigma} h_{\beta}{^\sigma} \nabla^\beta \xi^\alpha
      +
      2 \left(
          \delta \chi^\alpha_1 \nabla_\alpha \delta_1 R
          +
          2 \delta_1 R_{\alpha \beta} \nabla^\beta \delta \chi^\alpha_1
      \right)~.
\end{aligned}
\end{equation}
%===============================================
Note that the quantity $\Psi_{\rm R}$ is independent of the second-order perturbation of the GKV $\delta \chi_2^\mu$, and when $\mu=2$, it follows from Eq.~\ref{AKEPertII-KomarAKcond} that $\Psi_{\rm R}=0$. For vacuum spacetime, in the transverse-traceless gauge, at first order, one may choose $\delta \chi_1^\alpha$ to satisfy the background Killing equation, and hence, the terms dependent on $\delta \chi_1^\alpha$ disappear from $\Psi_{\rm R}$, so that $\Psi_{\rm R}$ depends only on the background quantities and the perturbation $h_{\mu \nu}$. It follows that to second order, $\Psi_R$ must be zero for vacuum perturbations (in which $\bar{R}_{\mu \nu}=0$) or any metric perturbation $h_{\mu \nu}$ which satisfies $h=\delta_1 R=0$ and permits a solution to the $\mu=2$ AKE.

If $\Psi_R=0$, the second-order perturbations should satisfy the following:
\begin{equation} \label{AKEPertII-KomarAKconstraint}
    \Box \Phi_2 = 0 ,
\end{equation}

\noindent where
\begin{equation} \label{Constraint2}
\Phi_2 := \left(\mu-2\right) \left\{
    \nabla \cdot \delta \chi_2
    -
    h^{\sigma \tau} \nabla_\alpha \left(\xi^\alpha h_{\sigma \tau}\right) \right\}.
\end{equation}

\noindent It follows that when $\Psi_R=0$, the field equations propagate the constraint $\Phi_2 = 0$ for the second-order perturbations (assuming initial data satisfying $\partial_t\Phi_2 = 0$). On a vacuum background, one may (assuming $\mu \neq 2$) use the constraint $\Phi_2 = 0$ to rewrite  Eq.~\ref{AKEPertII}
\begin{equation} \label{AKEPertIIrw}
\Box \delta \chi^{\nu}_2 - \nabla^\nu \left(\nabla \cdot \delta \chi_2 \right) = -j_2^\nu - m_2^\nu - (2 - \mu) \nabla^\nu \left\{ h^{\sigma \tau} \nabla_\alpha (\xi^\alpha h_{\sigma \tau}) \right\},
\end{equation}

\noindent so that the equation for $\delta \chi^{\nu}_2$ resembles the $\mu=2$ equation with a source. For second-order perturbations, the question of whether the perturbations suffer from dynamical instabilities due to the unboundedness of the Hamiltonian depends on the behavior of the rhs of Eq.~\ref{AKEPertIIrw}.

The claim that $\Psi_R = 0$ for transverse-traceless metric perturbations of vacuum spacetimes that admit Killing vectors suggests an identity for such perturbations. However, the arguments we have presented so far do not yet constitute a proof of such an identity, as they depend on the existence of solutions for the exact $\mu=2$ AKE. Though the AKE fails to admit a well-posed initial value problem for the $\mu=2$ parameter choice, there is some reason to expect that the failure is primarily due to nonuniqueness, rather than existence. For instance, it is straightforward to show that on vacuum spacetimes, the exact AKE becomes an identity for the gradient of an arbitrary function. Furthermore, one can show that in locally flat coordinates, the time derivatives for the time component of $\chi^\mu$ disappear, so that the AKE is an underdetermined dynamical system. One might therefore expect the existence of solutions to the exact $\mu=2$ AKE (and consequently, $\Phi_2 = 0$) to hold for a large class of transverse-traceless metric perturbations of vacuum spacetimes which admit Killing vectors.

%=======================================================================
%-----------------------------------------------------------------------
%
%		KOMAR CHARGE AND THERMODYNAMICAL INTERPRETATION
%
%-----------------------------------------------------------------------
%=======================================================================

\section{Perturbations of the Noether charge and its thermodynamical interpretation}\label{pake_thermodynamics}

In this section, we will discuss how the Noether charge associated with the GKV $\chi^\alpha$ associated with the AKE is affected by the perturbation of the metric. As we will demonstrate, the perturbed Noether charge will have an interesting thermodynamical interpretation. Applying Eq.~\ref{NoetherCurrent} to the solutions of the AKE, which correspond to the GKV field $\chi^{\mu}$, the Noether current for the GKVs in the perturbed spacetime takes the following form:
%-unc (see \ref{AppF} for a derivation)
%===============================================
\begin{align}\label{Noether_current_ake}
\bar{J}^{\mu}&=2\bar{R}{^\mu}_{\nu}\chi^{\nu}+(2-\bar{\mu})~\bar{\nabla}^{\mu}\left[\bar{\nabla}_{\sigma}\chi^{\sigma}\right]+\left(\bar{\nabla}_{\sigma}\chi^{\sigma}\right)\bar{g}^{\mu \alpha}\nabla_{\alpha}\left(2-\bar{\mu}\right)~,
\end{align}
%===============================================
where the AKE presented in Eq.~\ref{AlmostKillingEquation} has been used. Let us now use the fact, following Eq.~\ref{metric_pert}, that the spacetime metric can be expressed as $\bar{g}_{\mu \nu}=g_{\mu \nu}+h_{\mu \nu}$, where $h_{\mu \nu}$ is the perturbation, possibly due to some matter field entering the background spacetime geometry. As a consequence, we also have $\xi^{\sigma}\rightarrow \xi^{\sigma}+\delta \xi^{\sigma}=\chi^{\sigma}$, where $\chi^{\sigma}$ is the GKV, with the associated Noether current being given by Eq.~\ref{Noether_current_ake}.

Thus, the Noether current associated with the GKV field can be decomposed into the Komar current for the background Killing vector field $\xi^{\alpha}$ and a part containing additional corrections arising out of the gravitational perturbation $h_{\mu \nu}$ and the perturbation of the Killing vector field $\delta \xi^{\alpha}$. We then obtain the following expression for the Noether current associated with the GKV field $\chi^{\alpha}$:
%===============================================
\begin{align}
\bar{J}^{\nu}&=2R{^\nu}_{\sigma}\chi^{\sigma}+\left(-\square h^{\nu}_{\sigma}+R^{\nu \mu}h_{\mu \sigma}-R_{\sigma \mu}h^{\mu \nu}
-2R^{\nu}_{~ \mu \sigma \rho}h^{\mu \rho}\right)\chi^{\sigma}
\nonumber
\\
&+(2-\bar{\mu})\left(g^{\nu \alpha}-h^{\nu \alpha}\right)\nabla_{\alpha}\left(\nabla_{\sigma}\chi^{\sigma}+\frac{1}{2}\chi^{\sigma}\nabla_{\sigma}h\right)
+\left(g^{\nu \alpha}-h^{\nu \alpha}\right)\left(\nabla_{\sigma}\chi^{\sigma}+\frac{1}{2}\chi^{\sigma}\nabla_{\sigma}h\right)\nabla_{\alpha}\left(2-\bar{\mu}\right)~,
\end{align}
%===============================================
where we have used the Lorenz gauge condition to simplify the expression further. The above provides the expansion of the Noether current in terms of the gravitational perturbation; further expressing $\chi^{\sigma}=\xi^{\sigma}+\delta \xi^{\sigma}$, we obtain the following expression for the change in the Noether current:
%===============================================
\begin{align}\label{change_noether_current}
\delta J^{\nu}&=2R{^\nu}_{\sigma}\delta \xi^{\sigma}+\left(-\square h^{\nu}_{\sigma}+R^{\nu \mu}h_{\mu \sigma}-R_{\sigma \mu}h^{\mu \nu}-2R^{\nu}_{~ \mu \sigma \rho}h^{\mu \rho}\right)\left(\xi^{\sigma}+\delta \xi^{\sigma}\right)
\nonumber
\\
&+(2-\bar{\mu})\left(g^{\nu \alpha}-h^{\nu \alpha}\right)\nabla_{\alpha}\left(\nabla_{\sigma}\delta \xi^{\sigma}\right)
+\frac{1}{2}(2-\bar{\mu})g^{\nu \alpha}\nabla_{\alpha}\left[\left(\xi^{\sigma}+\delta \xi^{\sigma}\right)\nabla_{\sigma}h\right]
\nonumber
\\
&-\left[g^{\nu \alpha}\left(\nabla_{\sigma}\delta \xi^{\sigma}+\frac{1}{2}\left(\xi^{\sigma}+\delta \xi^{\sigma}\right)\nabla_{\sigma}h\right)-h^{\nu \alpha}\left(\nabla_{\sigma}\delta \xi^{\sigma}\right)\right]\left(\nabla_{\alpha}\delta \mu\right)~.
\end{align}
%===============================================
Here we have assumed that the background spacetime inherits Killing symmetry, and hence, $\xi^{\sigma}$ is a Killing vector field, such that $\nabla_{\sigma}\xi^{\sigma}=0$, which we have used in deriving the above expression. Using the expressions for the perturbations of the Ricci tensor and the Einstein tensor in the Lorenz gauge, the above change in the Noether current may be expressed in several ways, among which we quote the expression involving the Einstein tensor below
%-unc (see \ref{AppF} for the detailed derivation)
%===============================================
\begin{align}\label{change_noether_currentN}
\delta J^{\nu}&=2R{^\nu}_{\sigma}\delta \xi^{\sigma}+2\delta G^{\nu}_{\sigma}\left(\xi^{\sigma}+\delta \xi^{\sigma}\right)-\left(\frac{1}{2}\square h+R_{\mu \rho}h^{\mu \rho}\right)\left(\xi^{\nu}+\delta \xi^{\nu}\right)
\nonumber
\\
&\hskip 1 cm +(2-\bar{\mu})\left(g^{\nu \alpha}-h^{\nu \alpha}\right)\nabla_{\alpha}\left(\nabla_{\sigma}\delta \xi^{\sigma}\right)
+\frac{1}{2}(2-\bar{\mu})g^{\nu \alpha}\nabla_{\alpha}\left[\left(\xi^{\sigma}+\delta \xi^{\sigma}\right)\nabla_{\sigma}h\right]
\nonumber
\\
&\hskip 1 cm -\left[g^{\nu \alpha}\left(\nabla_{\sigma}\delta \xi^{\sigma}+\frac{1}{2}\left(\xi^{\sigma}+\delta \xi^{\sigma}\right)\nabla_{\sigma}h\right)-h^{\nu \alpha}\left(\nabla_{\sigma}\delta \xi^{\sigma}\right)\right]\left(\nabla_{\alpha}\delta \mu\right)~.
\end{align}
%===============================================
Observe that, using the perturbed Einstein equations, one can replace the perturbation of the Einstein tensor appearing in the above expression with the perturbation of the matter energy-momentum tensor. In that case, the object $\left(\delta G{^\nu}_{\sigma}\xi^{\sigma}\right)$ will correspond to the flux of the matter energy-momentum tensor through the Killing horizon, to which the Killing vector field is orthogonal. We will come back to this point later in this section, but we first discuss a couple of interesting limits:
%===============================================
%===============================================
%===============================================
\begin{itemize}

\item Even though we have treated the gravitational perturbation and the perturbation of the Killing vector field separately, the perturbation $\delta \xi^{\alpha}$ we are interested in is induced by the perturbation of the background spacetime. Thus it is natural to decompose the perturbed Killing vector field as $\delta \xi^{\mu}=\delta \chi^{\mu}_{1}+\delta \chi^{\mu}_{2}$ [cf. Eq.~\ref{GKVPert}], where $\delta \chi^{\mu}_{1}$ is linear in the gravitational perturbation and $\delta \chi^{\mu}_{2}$ is quadratic in the gravitational perturbation. Then to the linear order in the gravitational perturbation, the perturbed Noether current from  Eq.~\ref{change_noether_currentN} takes the following form:\footnote{One should be careful not to confuse this expression with  Eq.~\ref{KomarPAKE1}, which satisfies the identity $\nabla \cdot J_1 = 0$. $\delta J^\alpha$, on the other hand, satisfies $\bar{\nabla} \cdot (J+\delta J) = 0$.}
%===============================================
\begin{align}\label{change_noether_current_linearN}
\delta J^{\nu}&=2R{^\nu}_{\sigma}\delta \chi^{\sigma}_{1}+2\delta_1 R{^\nu}_{\sigma}\xi^{\sigma}
+(2-\mu)\nabla^{\nu}\left(\nabla \cdot \delta \chi_{1}\right)
+\frac{1}{2}(2-\mu)\nabla_{\nu}\left(\xi^{\sigma}\nabla_{\sigma}h\right)~
\\
&=2R{^\nu}_{\sigma}\delta \chi_{1}^{\sigma}+2\delta_1 G{^\nu}_{\sigma}\xi^{\sigma}-\left(\frac{1}{2}\square h+R_{\mu \rho}h^{\mu \rho}\right)\xi^{\nu}
%
% +(2-\mu)\nabla^{\nu}\left(\nabla \cdot \delta \chi_{1}\right)
% +\frac{1}{2}(2-\mu)\nabla_{\nu}\left(\xi^{\sigma}\nabla_{\sigma}h\right)~
+\frac{1}{2}\nabla_{\nu}\left(\xi^{\sigma}\nabla_{\sigma}h\right)~
\end{align}
%===============================================
where the Lorenz gauge condition has been used. In the first line of the above expression, we have expressed the change in the Noether current in terms of the change in the Ricci tensor. In the second line, we express the change in Noether current in terms of the change in the Einstein tensor, and apply the constraint $\Phi_1=0$ [with $\Phi_1$ given in Eq.~\ref{Constraint}], which follows from the fact that $\delta \chi^\mu_1$ satisfies Eq.~\ref{AKEPertI}, which as we showed earlier propagates the constraint $\Phi_1=0$ for an appropriate choice of initial data.

\item For spacetimes which may contain a dilute amount of nongravitational radiation on vacuum backgrounds, i.e., with $R_{\mu \nu}=0$, one can use the transverse-traceless gauge (effectively setting $h=0$), and hence, the above expression for the change in the Noether current simplifies considerably:
%===============================================
\begin{align}\label{reduction_nc_01}
\delta J^{\nu}&=2\delta_1 G{^\nu}_{\sigma}\xi^{\sigma}~.
\end{align}
%===============================================
Thus, using the perturbed Einstein equations, we find that given the constraint $\Phi_1=0$, the change in the Noether current to first order is simply equal to $16\pi(\delta T{^\nu}_{\sigma}\xi^{\sigma})$, which corresponds to the matter field flowing into the Killing horizon. This will have thermodynamical interpretation, as we compute the associated change in the Noether charge.

\item For spacetimes which contain a dilute amount of matter on vacuum backgrounds, one can instead employ the $\mu=2$ parameter choice, in which case the Noether current also simplifies to
%===============================================
\begin{align}\label{reduction_nc_02}
\delta J^{\nu}&=2\delta_1 R{^\nu}_{\sigma}\xi^{\sigma} = \left(2\delta_1 G{^\nu}_{\sigma} + \delta^{\nu}_{\sigma}~ \delta_1 R \right)\xi^{\sigma} ~.
\end{align}
%===============================================
We see that for matter fields, the Noether current does not measure energy and momentum in the sense of the energy-momentum tensor, due to the term containing $\delta_1 R$. However, one may nonetheless still regard the Noether current and its associated charge as a measure of the matter content, and an analysis \cite{Lynden-BellBicak2017} comparing Komar integrals for radiation with that of matter in cylindrical symmetry suggests that charges constructed from Noether currents measure the effective gravitating mass.

\end{itemize}
%===============================================
%===============================================
%===============================================
Thus, we have discussed situations of physical interest and how the change in the Noether current can be simplified and interpreted in these scenarios. We will now proceed to compute the change in the Noether charge due to the gravitational perturbation and the perturbation of the Killing vector field.

In order to determine the Noether charge, we have to integrate the Noether current over a three-surface. One may define such a hypersurface as a level surface of some scalar function $\Phi=\Phi(x)$; defined in this way, the surface is held fixed in the manifold and does not change under metric perturbations. However, the measure on the surface will change. In particular, for a $\Phi=\textrm{constant}$ surface, the integration measure is $d\Sigma_{\alpha}=d^{3}x\sqrt{h}n_{\alpha}$, where $n_{\alpha}$ is the normalized normal on this surface, and $h$ is the determinant of the induced metric on the $\Phi=\textrm{constant}$ surface. It is also possible to express the integration measure on a $\Phi=\textrm{constant}$ surface as $d\Sigma_{\alpha}=d^{3}x\sqrt{-g}\nabla_{\alpha}\Phi$, where $\nabla_{\alpha}\Phi$ is the unnormalized normal to the $\Phi=\textrm{constant}$ surface. Under metric perturbations, the measure changes as $\sqrt{-g}\rightarrow \sqrt{-\bar{g}}=\sqrt{-g}\{1+(1/2)h\}$. Thus, the Noether charge associated with the GKV field $\chi^{\alpha}=\xi^{\alpha}+\delta \xi^{\alpha}$ becomes
%===============================================
\begin{align}
\bar{Q}[\xi^\cdot+\delta \xi^\cdot]&=\int d\bar{\Sigma}_{\alpha}\bar{J}^{\alpha}=\int d\Sigma_{\alpha}\left(1+\frac{1}{2}h\right)\left(J^{\alpha}+\delta J^{\alpha}\right)
\nonumber
\\
&=Q[\xi^\cdot]+\int d\Sigma_{\alpha}\left(\frac{1}{2}h\right)J^{\alpha}+\int d\Sigma_{\alpha}\delta J^{\alpha}
+\int d\Sigma_{\alpha}\left(\frac{1}{2}h\right)\delta J^{\alpha}~,
\end{align}
%===============================================
where $Q[\xi^\cdot]$ denotes the Komar charge of the background Killing vector field. It is straightforward to read off the change in the Noether charge, which takes the following form:
%===============================================
\begin{align}
\delta Q\equiv \bar{Q}[\xi^\cdot+\delta \xi^\cdot]-Q[\xi^\cdot]&=\int d\Sigma_{\alpha}\left(\frac{1}{2}h\right)J^{\alpha}+\int d\Sigma_{\alpha}\delta J^{\alpha}+\int d\Sigma_{\alpha}\left(\frac{1}{2}h\right)\delta J^{\alpha}~.
\end{align}
%===============================================
Note that the first term is due to the perturbation of the integration measure, while the second and third terms result from a change in the Noether current. As in the case of the Noether current, this expression can be simplified further if we assume that the perturbations correspond to dilute radiation in vacuum background spacetime. The transverse-traceless gauge condition can then be imposed, and thus, the trace of the perturbation $h$ can be set to zero. Therefore, the above expression for the change in the Noether charge becomes
%-unc (for a derivation, see \ref{AppF})
%===============================================
\begin{align}
\delta Q=2\int d\Sigma_{\alpha}\delta G{^\alpha}_{\beta}\xi^{\beta}~.
\end{align}
%===============================================
As we will discuss below, this expression can be understood from a physical as well as thermodynamical perspective.

We consider the case where the metric $g_{\mu \nu}$ is a vacuum black hole spacetime, with a Killing vector field $\xi^{\alpha}$ defining the Killing horizon, to which the Killing vector is orthogonal. Since the Killing horizon is a null surface, it follows that we can consider the surface on which the Noether charge is computed to be the Killing horizon, which is bound to have a thermodynamic interpretation. In particular, for a generic null surface, the Noether charge associated with a Killing-like vector corresponds to $Q=16\pi ~\mathcal{T} S$, where $\mathcal{T}$ is the temperature associated with the null surface and $S$ is the associated entropy, which for general relativity is simply (area/4) \cite{Chakraborty:2015hna,Chakraborty:2019doh}. As evident from the Vaidya solution considered in \ref{AKE_review} and
%-unc also from the general calculation presented in \ref{AppF}
it follows that to first order, under the assumptions considered here, the temperature does not change, and hence, the above change in the Noether charge may be interpreted as
%===============================================
\begin{align}
\mathcal{T} \delta S=\int d\Sigma_{\alpha}\delta T{^\alpha}_{\beta}\xi^{\beta},
\end{align}
%===============================================
which is the Clausius relation. In brief, this suggests that due to radiation falling into the Killing horizon, the spacetime is perturbed, and the Killing vector also ceases to be Killing, rather it becomes a GKV. This perturbation under the appropriate limit yields the Clausius relation. Thus, the formalism developed here for the case of radiation provides a close correspondence with the thermodynamic nature of gravity. One may make a similar argument for the case of nonradiative matter with the $\mu=2$ parameter choice for null surfaces, provided that $d\Sigma_\alpha \xi^\alpha = 0$, or by interpreting the Noether charge in terms of gravitating mass, as suggested in \cite{Lynden-BellBicak2017}.

%=======================================================================
%-----------------------------------------------------------------------
%
%		FINAL REMARKS
%
%-----------------------------------------------------------------------
%=======================================================================
\section{Final Remarks}\label{pake_conclusions}

Killing vectors play a central role in characterizing spacetime symmetries, which are crucial in determining the conserved quantities that can be constructed in a given spacetime geometry. Systems of astrophysical interest are often symmetric only in a first approximation, and the spacetime geometries for such systems are often more accurately described in terms of a symmetric spacetime background with perturbations (the latter due to gravitational radiation or inhomogeneities and highly dynamical behavior in matter fields) which explicitly break the symmetry of the background. This motivates the perturbative study of the AKE, the solutions of which provide the generalizations of Killing vectors (which we refer to as GKVs) appropriate for the perturbed spacetime.

In this article, we have examined in detail the construction and behavior of these GKVs as perturbative solutions to the AKE associated with the metric perturbations of vacuum and nonvacuum spacetimes, which admit a Killing vector field. This has been achieved in two steps: (a) by considering the perturbation of the action yielding the AKE and then varying the same with respect to the perturbed GKV and (b) by perturbing the AKE and hence determining the evolution equation of the perturbed GKV. The matching of both of these equations explicitly demonstrates the internal consistency of these results.

Additionally, it turns out that the hyperbolicity and Hamiltonian stability of the perturbed GKV remains unchanged compared to its unperturbed counterpart if the GKV and the metric perturbations are kept independent. However, we have found that in the case where the perturbations of the GKV are sourced by the metric perturbation, the problem of an unbounded Hamiltonian can be avoided at first order, and at second order, the problem may also be avoided if the metric perturbations are transverse and traceless (assuming the perturbations remain well behaved), and perturbative solutions to the $\mu=2$ AKE exist to second order. We have found that the first-order equations trivialize (they reduce to the background AKE) for traceless metric perturbations; for dilute radiation, the second-order case is necessary. We have also examined the first-order behavior of the Noether current constructed from a GKV and its associated charge. Intriguingly, it turns out that the conservation of the Noether current introduces additional constraints in the theory, which helps significantly to simplify the evolution equation for the perturbed GKV. In particular, if the perturbed matter energy-momentum tensor is traceless, i.e., the perturbation is due to null matter field, it follows that the first-order perturbation of the GKV can be absorbed within the background Killing vector field. To second order, we find that the second-order perturbation always yields nontrivial modifications to the background Killing vector field.

Finally, the perturbation of a background spacetime respecting Killing symmetry also has interesting thermodynamic implications. In particular, as we have demonstrated, the perturbation of the Noether charge to first order can be expressed as $\mathcal{T} \delta S$. This is because to first order in the perturbation, under these assumptions, the surface gravity does not change. This is also apparent from the example of Vaidya spacetime considered in this work, which further corroborates our claims regarding the thermodynamic interpretation for the perturbed Noether charge and currents associated with the generalized Killing vector fields.

%=======================================================================

%-----------------------------------------------------------------------
%-----------------------------------
%-----------------
%--------
%---
%-
%
%
%-
%---
%--------
%-----------------
%-----------------------------------
%-----------------------------------------------------------------------

%=======================================================================
%		ACKNOWLEDGMENTS
%=======================================================================

\begin{acknowledgments}

We are grateful to Edgar Gasperin and David Hilditch for helpful discussions. Some of the calculations were performed using the xAct package \cite{xActbib} for \textit{Mathematica}. J. C. F. acknowledges support from FCT Grants No. PTDC/MAT-APL/30043/2017 and No. UIDB/00099/2020. Research of S. C. is funded by the INSPIRE Faculty fellowship from DST, Government of India (Reg. No. DST/INSPIRE/04/2018/000893) and by the Start-Up Research Grant from SERB, DST, Government of India (Reg. No. SRG/2020/000409).

\end{acknowledgments}

%=======================================================================
%-----------------------------------------------------------------------
%
%		APPENDICES
%
%-----------------------------------------------------------------------
%=======================================================================
\appendix
\labelformat{section}{Appendix #1}
\labelformat{subsection}{Appendix #1}

%===============================================
%-----------------------------------------------
%
%		Action for AKE
%
%-----------------------------------------------
%===============================================
\section{Variational principle for AKE} \label{AppA}
Here, we review the derivation of the AKE from a variational principle, following the notation and conventions of
Sec. III A.
%   \ref{SSNotConv}.
We rewrite here the action given in Eq.~\ref{ActionComp} in expanded form:
%===============================================
\begin{align}\label{action_app_killing}
\mathcal{A}[\chi^\cdot]=\int_{\mathcal{V}} d^{4}x\sqrt{-\bg}\left[-\frac{1}{4}\left(\bd_{\mu}\chi_{\nu}+\bd_{\nu}\chi_{\mu}\right)\left(\bd^{\mu}\chi^{\nu}+\bd^{\nu}\chi^{\mu}\right)+\frac{\bar{\mu}}{2}\left(\bd_{\mu}\chi^{\mu}\right)^{2}\right]~,
\end{align}
%===============================================
where $\bar{g}_{\alpha \beta}$ is the metric of the perturbed spacetime, and $\bar{\mu}\equiv \bar{\mu}(x)$ is an arbitrary function of the spacetime coordinates. Varying the above action with respect to arbitrary variations of $\chi^{\mu}$, including endpoint contributions, we obtain
%===============================================
\begin{align}
\Delta \mathcal{A}&=\int_{\mathcal{V}} d^{4}x\sqrt{-\bg}\Big[-\bg_{\mu \alpha}\bg_{\nu \beta}\left(\bd^{\mu}\chi^{\nu}+\bd^{\nu}\chi^{\mu}\right)\left(\bd^{\alpha}\Delta \chi^{\beta}\right) +\bar{\mu}\left(\bd_{\mu}\chi^{\mu}\right)\left(\bd_{\alpha}\Delta\chi^{\alpha}\right)\Big]
\nonumber
\\
&\hskip 1 cm +\int _{\partial \mathcal{V}}d^{3}x~\sqrt{-\bg}\left[-\frac{1}{4}\left(\bd_{\mu}\chi_{\nu}+\bd_{\nu}\chi_{\mu}\right)\left(\bd^{\mu}\chi^{\nu}+\bd^{\nu}\chi^{\mu}\right)+\frac{\bar{\mu}}{2}\left(\bd_{\mu}\chi^{\mu}\right)^{2}\right]\Delta x^{\alpha}\bd_{\alpha}\phi~,
\end{align}
%===============================================
where $\partial \mathcal{V}$ is the boundary surface of the full spacetime volume $\mathcal{V}$ described by some arbitrary scalar function, $\phi(x)=\textrm{constant}$. By performing integration by parts, the above expression for the variation of the action can be further simplified and it will yield several boundary terms. Since these boundary terms will not play any significant role immediately, we will neglect all the boundary contributions, and hence, the above variation of the action functional can be expressed in the following manner:
%===============================================
\begin{align}
\Delta \mathcal{A}&=\int_{\mathcal{V}} d^{4}x\sqrt{-\bg}\Big[\Delta \chi^{\beta}\bg_{\nu \beta}\left(\bar{\square} \chi^{\nu}+[\bd_{\mu},\bd^{\nu}]\chi^{\mu}+\bd^{\nu}\bd_{\mu}\chi^{\mu}\right)-\Delta\chi^{\beta}\bd_{\beta}\left(\bar{\mu}~\bd_{\sigma}\chi^{\sigma}\right)\Big]
\nonumber
\\
&=\int_{\mathcal{V}} d^{4}x\sqrt{-\bg}~\bg_{\nu \beta}~\Delta \chi^{\beta}\Big[\bar{\square} \chi^{\nu}+\bar{R}{^\nu}_{\mu}\chi^{\mu}+\bd^{\nu}\left\{\left(1-\bar{\mu}\right)\bd_{\sigma}\chi^{\sigma}\right\}\Big]~.
\end{align}
%===============================================
Here we have used the fact that the commutator of covariant derivatives acting on a vector is given by the Riemann tensor. Thus, setting the variation of the action functional $\Delta \mathcal{A}$ to be zero, for arbitrary variations of the GKV field $\chi^{\beta}$, we obtain
%===============================================
\begin{align}\label{pert_ake_var}
\sqrt{-\bg}~\bg_{\nu \beta}\Big[\bar{\square} \chi^{\nu}+\bar{R}^{\nu}_{\mu}\chi^{\mu}+\bd^{\nu}\left\{\left(1-\bar{\mu}\right)\bd_{\sigma}\chi^{\sigma}\right\}\Big]=0~.
\end{align}
%===============================================
Since we are interested in nondegenerate spacetime, i.e., spacetimes with a metric $\bar{g}_{\alpha \beta}$ with nonzero determinant and nontrivial inverse, the above equation can be casted in the following form:
%===============================================
\begin{align}\label{pert_ake}
\bar{\square} \chi^{\nu}+\bar{R}{^\nu}_{\mu}\chi^{\mu}+\bd^{\nu}\left\{\left(1-\bar{\mu}\right)\bd_{\sigma}\chi^{\sigma}\right\}=0~.
\end{align}
%===============================================
The above equation corresponds to the AKE satisfied by the GKV $\chi^{\mu}$ in the exact spacetime, with metric $\bar{g}_{\mu \nu}$. Note that, in the above expression we have kept $\bar{\mu}$ inside the derivative terms since it is a function of the spacetime coordinates.

\bibliography{PAKE}

%merlin.mbs apsrev4-1.bst 2010-07-25 4.21a (PWD, AO, DPC) hacked
%Control: key (0)
%Control: author (72) initials jnrlst
%Control: editor formatted (1) identically to author
%Control: production of article title (-1) disabled
%Control: page (0) single
%Control: year (1) truncated
%Control: production of eprint (0) enabled
\begin{thebibliography}{44}%
\makeatletter
\providecommand \@ifxundefined [1]{%
 \@ifx{#1\undefined}
}%
\providecommand \@ifnum [1]{%
 \ifnum #1\expandafter \@firstoftwo
 \else \expandafter \@secondoftwo
 \fi
}%
\providecommand \@ifx [1]{%
 \ifx #1\expandafter \@firstoftwo
 \else \expandafter \@secondoftwo
 \fi
}%
\providecommand \natexlab [1]{#1}%
\providecommand \enquote  [1]{``#1''}%
\providecommand \bibnamefont  [1]{#1}%
\providecommand \bibfnamefont [1]{#1}%
\providecommand \citenamefont [1]{#1}%
\providecommand \href@noop [0]{\@secondoftwo}%
\providecommand \href [0]{\begingroup \@sanitize@url \@href}%
\providecommand \@href[1]{\@@startlink{#1}\@@href}%
\providecommand \@@href[1]{\endgroup#1\@@endlink}%
\providecommand \@sanitize@url [0]{\catcode `\\12\catcode `\$12\catcode
  `\&12\catcode `\#12\catcode `\^12\catcode `\_12\catcode `\%12\relax}%
\providecommand \@@startlink[1]{}%
\providecommand \@@endlink[0]{}%
\providecommand \url  [0]{\begingroup\@sanitize@url \@url }%
\providecommand \@url [1]{\endgroup\@href {#1}{\urlprefix }}%
\providecommand \urlprefix  [0]{URL }%
\providecommand \Eprint [0]{\href }%
\providecommand \doibase [0]{http://dx.doi.org/}%
\providecommand \selectlanguage [0]{\@gobble}%
\providecommand \bibinfo  [0]{\@secondoftwo}%
\providecommand \bibfield  [0]{\@secondoftwo}%
\providecommand \translation [1]{[#1]}%
\providecommand \BibitemOpen [0]{}%
\providecommand \bibitemStop [0]{}%
\providecommand \bibitemNoStop [0]{.\EOS\space}%
\providecommand \EOS [0]{\spacefactor3000\relax}%
\providecommand \BibitemShut  [1]{\csname bibitem#1\endcsname}%
\let\auto@bib@innerbib\@empty
%</preamble>
\bibitem [{\citenamefont {Matzner}(1968)}]{Matzner1968}%
  \BibitemOpen
  \bibfield  {author} {\bibinfo {author} {\bibfnamefont {R.~A.}\ \bibnamefont
  {Matzner}},\ }\href {\doibase 10.1063/1.1664495} {\bibfield  {journal}
  {\bibinfo  {journal} {J. Math. Phys. (N.Y.)}\ }\textbf {\bibinfo {volume}
  {9}},\ \bibinfo {pages} {1657} (\bibinfo {year} {1968})}\BibitemShut
  {NoStop}%
\bibitem [{\citenamefont {Beetle}(2008)}]{Beetle2008}%
  \BibitemOpen
  \bibfield  {author} {\bibinfo {author} {\bibfnamefont {C.}~\bibnamefont
  {Beetle}},\ }\href@noop {} {\  (\bibinfo {year} {2008})},\ \Eprint
  {http://arxiv.org/abs/0808.1745} {arXiv:0808.1745 [gr-qc]} \BibitemShut
  {NoStop}%
\bibitem [{\citenamefont {Beetle}\ and\ \citenamefont
  {Wilder}(2014)}]{BeetleWilder2014}%
  \BibitemOpen
  \bibfield  {author} {\bibinfo {author} {\bibfnamefont {C.}~\bibnamefont
  {Beetle}}\ and\ \bibinfo {author} {\bibfnamefont {S.}~\bibnamefont
  {Wilder}},\ }\href {http://stacks.iop.org/0264-9381/31/i=7/a=075009}
  {\bibfield  {journal} {\bibinfo  {journal} {Classical Quantum Gravity}\
  }\textbf {\bibinfo {volume} {31}},\ \bibinfo {pages} {075009} (\bibinfo
  {year} {2014})},\ \Eprint {http://arxiv.org/abs/1401.0074} {arXiv:1401.0074
  [gr-qc]} \BibitemShut {NoStop}%
\bibitem [{\citenamefont {Brunekreef}\ and\ \citenamefont
  {Reitz}(2020)}]{BrunekreefReitz2020}%
  \BibitemOpen
  \bibfield  {author} {\bibinfo {author} {\bibfnamefont {J.}~\bibnamefont
  {Brunekreef}}\ and\ \bibinfo {author} {\bibfnamefont {M.}~\bibnamefont
  {Reitz}},\ }\href@noop {} {\  (\bibinfo {year} {2020})},\ \Eprint
  {http://arxiv.org/abs/2012.14518} {arXiv:2012.14518 [gr-qc]} \BibitemShut
  {NoStop}%
\bibitem [{\citenamefont {Garfinkle}\ and\ \citenamefont
  {Gundlach}(1999)}]{GarfinkleGundlach1999}%
  \BibitemOpen
  \bibfield  {author} {\bibinfo {author} {\bibfnamefont {D.}~\bibnamefont
  {Garfinkle}}\ and\ \bibinfo {author} {\bibfnamefont {C.}~\bibnamefont
  {Gundlach}},\ }\href {\doibase 10.1088/0264-9381/16/12/325} {\bibfield
  {journal} {\bibinfo  {journal} {Classical Quantum Gravity}\ }\textbf
  {\bibinfo {volume} {16}},\ \bibinfo {pages} {4111} (\bibinfo {year}
  {1999})},\ \Eprint {http://arxiv.org/abs/gr-qc/9908016} {arXiv:gr-qc/9908016}
  \BibitemShut {NoStop}%
\bibitem [{\citenamefont {Harte}(2008)}]{Harte2008}%
  \BibitemOpen
  \bibfield  {author} {\bibinfo {author} {\bibfnamefont {A.~I.}\ \bibnamefont
  {Harte}},\ }\href {http://stacks.iop.org/0264-9381/25/i=20/a=205008}
  {\bibfield  {journal} {\bibinfo  {journal} {Classical Quantum Gravity}\
  }\textbf {\bibinfo {volume} {25}},\ \bibinfo {pages} {205008} (\bibinfo
  {year} {2008})},\ \Eprint {http://arxiv.org/abs/0805.4259} {arXiv:0805.4259
  [gr-qc]} \BibitemShut {NoStop}%
\bibitem [{\citenamefont {Taubes}(1978)}]{Taubes1978}%
  \BibitemOpen
  \bibfield  {author} {\bibinfo {author} {\bibfnamefont {C.~H.}\ \bibnamefont
  {Taubes}},\ }\href {\doibase 10.1063/1.523859} {\bibfield  {journal}
  {\bibinfo  {journal} {J. Math. Phys. (N.Y.)}\ }\textbf {\bibinfo {volume}
  {19}},\ \bibinfo {pages} {1515} (\bibinfo {year} {1978})}\BibitemShut
  {NoStop}%
\bibitem [{\citenamefont {Bona}\ \emph {et~al.}(2005)\citenamefont {Bona},
  \citenamefont {Carot},\ and\ \citenamefont
  {Palenzuela-Luque}}]{Bonaetal2005}%
  \BibitemOpen
  \bibfield  {author} {\bibinfo {author} {\bibfnamefont {C.}~\bibnamefont
  {Bona}}, \bibinfo {author} {\bibfnamefont {J.}~\bibnamefont {Carot}}, \ and\
  \bibinfo {author} {\bibfnamefont {C.}~\bibnamefont {Palenzuela-Luque}},\
  }\href {\doibase 10.1103/PhysRevD.72.124010} {\bibfield  {journal} {\bibinfo
  {journal} {Phys. Rev. D}\ }\textbf {\bibinfo {volume} {72}},\ \bibinfo
  {pages} {124010} (\bibinfo {year} {2005})},\ \Eprint
  {http://arxiv.org/abs/gr-qc/0509015} {arXiv:gr-qc/0509015} \BibitemShut
  {NoStop}%
\bibitem [{\citenamefont {Komar}(1959)}]{Komar1959}%
  \BibitemOpen
  \bibfield  {author} {\bibinfo {author} {\bibfnamefont {A.}~\bibnamefont
  {Komar}},\ }\href {\doibase 10.1103/PhysRev.113.934} {\bibfield  {journal}
  {\bibinfo  {journal} {Phys. Rev.}\ }\textbf {\bibinfo {volume} {113}},\
  \bibinfo {pages} {934} (\bibinfo {year} {1959})}\BibitemShut {NoStop}%
\bibitem [{\citenamefont {Komar}(1962)}]{Komar1962}%
  \BibitemOpen
  \bibfield  {author} {\bibinfo {author} {\bibfnamefont {A.}~\bibnamefont
  {Komar}},\ }\href {\doibase 10.1103/PhysRev.127.1411} {\bibfield  {journal}
  {\bibinfo  {journal} {Phys. Rev.}\ }\textbf {\bibinfo {volume} {127}},\
  \bibinfo {pages} {1411} (\bibinfo {year} {1962})}\BibitemShut {NoStop}%
\bibitem [{\citenamefont {Ruiz}\ \emph {et~al.}(2014)\citenamefont {Ruiz},
  \citenamefont {Palenzuela},\ and\ \citenamefont {Bona}}]{Ruizetal2014}%
  \BibitemOpen
  \bibfield  {author} {\bibinfo {author} {\bibfnamefont {M.}~\bibnamefont
  {Ruiz}}, \bibinfo {author} {\bibfnamefont {C.}~\bibnamefont {Palenzuela}}, \
  and\ \bibinfo {author} {\bibfnamefont {C.}~\bibnamefont {Bona}},\ }\href
  {\doibase 10.1103/PhysRevD.89.025011} {\bibfield  {journal} {\bibinfo
  {journal} {Phys. Rev. D}\ }\textbf {\bibinfo {volume} {89}},\ \bibinfo
  {pages} {025011} (\bibinfo {year} {2014})},\ \Eprint
  {http://arxiv.org/abs/1312.3466} {arXiv:1312.3466 [gr-qc]} \BibitemShut
  {NoStop}%
\bibitem [{\citenamefont {Feng}(2018)}]{FengCurrents2018}%
  \BibitemOpen
  \bibfield  {author} {\bibinfo {author} {\bibfnamefont {J.~C.}\ \bibnamefont
  {Feng}},\ }\href {\doibase 10.1103/PhysRevD.98.104035} {\bibfield  {journal}
  {\bibinfo  {journal} {Phys. Rev. D}\ }\textbf {\bibinfo {volume} {98}},\
  \bibinfo {pages} {104035} (\bibinfo {year} {2018})},\ \Eprint
  {http://arxiv.org/abs/1811.05312} {arXiv:1811.05312 [gr-qc]} \BibitemShut
  {NoStop}%
\bibitem [{\citenamefont {Peng}(2020)}]{Peng2020}%
  \BibitemOpen
  \bibfield  {author} {\bibinfo {author} {\bibfnamefont {J.-J.}\ \bibnamefont
  {Peng}},\ }\href {\doibase 10.1088/1572-9494/ab8a14} {\bibfield  {journal}
  {\bibinfo  {journal} {Communications in Theoretical Physics}\ }\textbf
  {\bibinfo {volume} {72}},\ \bibinfo {pages} {065402} (\bibinfo {year}
  {2020})},\ \Eprint {http://arxiv.org/abs/1909.09921} {arXiv:1909.09921
  [gr-qc]} \BibitemShut {NoStop}%
\bibitem [{\citenamefont {Peng}\ and\ \citenamefont {Zou}(2019)}]{Peng2019}%
  \BibitemOpen
  \bibfield  {author} {\bibinfo {author} {\bibfnamefont {J.-J.}\ \bibnamefont
  {Peng}}\ and\ \bibinfo {author} {\bibfnamefont {C.-L.}\ \bibnamefont {Zou}},\
  }\href@noop {} {\  (\bibinfo {year} {2019})},\ \Eprint
  {http://arxiv.org/abs/1910.00339} {arXiv:1910.00339 [gr-qc]} \BibitemShut
  {NoStop}%
\bibitem [{\citenamefont {Feng}\ \emph {et~al.}(2019)\citenamefont {Feng},
  \citenamefont {Gasper\'{\i}n},\ and\ \citenamefont {Williams}}]{FGW2019}%
  \BibitemOpen
  \bibfield  {author} {\bibinfo {author} {\bibfnamefont {J.~C.}\ \bibnamefont
  {Feng}}, \bibinfo {author} {\bibfnamefont {E.}~\bibnamefont {Gasper\'{\i}n}},
  \ and\ \bibinfo {author} {\bibfnamefont {J.~L.}\ \bibnamefont {Williams}},\
  }\href {\doibase 10.1103/PhysRevD.100.124034} {\bibfield  {journal} {\bibinfo
   {journal} {Phys. Rev. D}\ }\textbf {\bibinfo {volume} {100}},\ \bibinfo
  {pages} {124034} (\bibinfo {year} {2019})},\ \Eprint
  {http://arxiv.org/abs/1911.04354} {arXiv:1911.04354 [gr-qc]} \BibitemShut
  {NoStop}%
\bibitem [{\citenamefont {Katz}(1985)}]{Katz1985}%
  \BibitemOpen
  \bibfield  {author} {\bibinfo {author} {\bibfnamefont {J.}~\bibnamefont
  {Katz}},\ }\href {http://stacks.iop.org/0264-9381/2/i=3/a=018} {\bibfield
  {journal} {\bibinfo  {journal} {Classical Quantum Gravity}\ }\textbf
  {\bibinfo {volume} {2}},\ \bibinfo {pages} {423} (\bibinfo {year}
  {1985})}\BibitemShut {NoStop}%
\bibitem [{\citenamefont {Bak}\ \emph {et~al.}(1994)\citenamefont {Bak},
  \citenamefont {Cangemi},\ and\ \citenamefont {Jackiw}}]{Baketal1993}%
  \BibitemOpen
  \bibfield  {author} {\bibinfo {author} {\bibfnamefont {D.}~\bibnamefont
  {Bak}}, \bibinfo {author} {\bibfnamefont {D.}~\bibnamefont {Cangemi}}, \ and\
  \bibinfo {author} {\bibfnamefont {R.}~\bibnamefont {Jackiw}},\ }\href
  {\doibase 10.1103/PhysRevD.49.5173} {\bibfield  {journal} {\bibinfo
  {journal} {Phys. Rev. D}\ }\textbf {\bibinfo {volume} {49}},\ \bibinfo
  {pages} {5173} (\bibinfo {year} {1994})},\ \Eprint
  {http://arxiv.org/abs/hep-th/9310025} {arXiv:hep-th/9310025} \BibitemShut
  {NoStop}%
\bibitem [{\citenamefont {Katz}\ \emph {et~al.}(1997)\citenamefont {Katz},
  \citenamefont {Bi\ifmmode~\check{c}\else \v{c}\fi{}\'ak},\ and\ \citenamefont
  {Lynden-Bell}}]{KBL1997}%
  \BibitemOpen
  \bibfield  {author} {\bibinfo {author} {\bibfnamefont {J.}~\bibnamefont
  {Katz}}, \bibinfo {author} {\bibfnamefont {J.}~\bibnamefont
  {Bi\ifmmode~\check{c}\else \v{c}\fi{}\'ak}}, \ and\ \bibinfo {author}
  {\bibfnamefont {D.}~\bibnamefont {Lynden-Bell}},\ }\href {\doibase
  10.1103/PhysRevD.55.5957} {\bibfield  {journal} {\bibinfo  {journal} {Phys.
  Rev. D}\ }\textbf {\bibinfo {volume} {55}},\ \bibinfo {pages} {5957}
  (\bibinfo {year} {1997})}\BibitemShut {NoStop}%
\bibitem [{\citenamefont {Obukhov}\ and\ \citenamefont
  {Rubilar}(2006{\natexlab{a}})}]{Obukhovetal2006a}%
  \BibitemOpen
  \bibfield  {author} {\bibinfo {author} {\bibfnamefont {Y.~N.}\ \bibnamefont
  {Obukhov}}\ and\ \bibinfo {author} {\bibfnamefont {G.~F.}\ \bibnamefont
  {Rubilar}},\ }\href {\doibase 10.1103/PhysRevD.73.124017} {\bibfield
  {journal} {\bibinfo  {journal} {Phys. Rev. D}\ }\textbf {\bibinfo {volume}
  {73}},\ \bibinfo {pages} {124017} (\bibinfo {year} {2006}{\natexlab{a}})},\
  \Eprint {http://arxiv.org/abs/gr-qc/0605045} {arXiv:gr-qc/0605045}
  \BibitemShut {NoStop}%
\bibitem [{\citenamefont {Obukhov}\ and\ \citenamefont
  {Rubilar}(2006{\natexlab{b}})}]{Obukhovetal2006b}%
  \BibitemOpen
  \bibfield  {author} {\bibinfo {author} {\bibfnamefont {Y.~N.}\ \bibnamefont
  {Obukhov}}\ and\ \bibinfo {author} {\bibfnamefont {G.~F.}\ \bibnamefont
  {Rubilar}},\ }\href {\doibase 10.1103/PhysRevD.74.064002} {\bibfield
  {journal} {\bibinfo  {journal} {Phys. Rev. D}\ }\textbf {\bibinfo {volume}
  {74}},\ \bibinfo {pages} {064002} (\bibinfo {year} {2006}{\natexlab{b}})},\
  \Eprint {http://arxiv.org/abs/gr-qc/0608064} {arXiv:gr-qc/0608064}
  \BibitemShut {NoStop}%
\bibitem [{\citenamefont {Deruelle}\ \emph {et~al.}(2004)\citenamefont
  {Deruelle}, \citenamefont {Katz},\ and\ \citenamefont
  {Ogushi}}]{Deruelleetal2004}%
  \BibitemOpen
  \bibfield  {author} {\bibinfo {author} {\bibfnamefont {N.}~\bibnamefont
  {Deruelle}}, \bibinfo {author} {\bibfnamefont {J.}~\bibnamefont {Katz}}, \
  and\ \bibinfo {author} {\bibfnamefont {S.}~\bibnamefont {Ogushi}},\ }\href
  {http://stacks.iop.org/0264-9381/21/i=8/a=004} {\bibfield  {journal}
  {\bibinfo  {journal} {Classical Quantum Gravity}\ }\textbf {\bibinfo {volume}
  {21}},\ \bibinfo {pages} {1971} (\bibinfo {year} {2004})},\ \Eprint
  {http://arxiv.org/abs/gr-qc/0310098} {arXiv:gr-qc/0310098} \BibitemShut
  {NoStop}%
\bibitem [{\citenamefont {Obukhov}\ \emph {et~al.}(2006)\citenamefont
  {Obukhov}, \citenamefont {Rubilar},\ and\ \citenamefont
  {Pereira}}]{Obukhovetal2006c}%
  \BibitemOpen
  \bibfield  {author} {\bibinfo {author} {\bibfnamefont {Y.~N.}\ \bibnamefont
  {Obukhov}}, \bibinfo {author} {\bibfnamefont {G.~F.}\ \bibnamefont
  {Rubilar}}, \ and\ \bibinfo {author} {\bibfnamefont {J.~G.}\ \bibnamefont
  {Pereira}},\ }\href {\doibase 10.1103/PhysRevD.74.104007} {\bibfield
  {journal} {\bibinfo  {journal} {Phys. Rev. D}\ }\textbf {\bibinfo {volume}
  {74}},\ \bibinfo {pages} {104007} (\bibinfo {year} {2006})},\ \Eprint
  {http://arxiv.org/abs/gr-qc/0610092} {arXiv:gr-qc/0610092} \BibitemShut
  {NoStop}%
\bibitem [{\citenamefont {Obukhov}\ and\ \citenamefont
  {Puetzfeld}(2013)}]{Obukhovetal2013}%
  \BibitemOpen
  \bibfield  {author} {\bibinfo {author} {\bibfnamefont {Y.~N.}\ \bibnamefont
  {Obukhov}}\ and\ \bibinfo {author} {\bibfnamefont {D.}~\bibnamefont
  {Puetzfeld}},\ }\href {\doibase 10.1103/PhysRevD.87.081502} {\bibfield
  {journal} {\bibinfo  {journal} {Phys. Rev. D}\ }\textbf {\bibinfo {volume}
  {87}},\ \bibinfo {pages} {081502} (\bibinfo {year} {2013})},\ \Eprint
  {http://arxiv.org/abs/1303.6050} {arXiv:1303.6050 [gr-qc]} \BibitemShut
  {NoStop}%
\bibitem [{\citenamefont {Obukhov}\ and\ \citenamefont
  {Puetzfeld}(2014)}]{Obukhovetal2014}%
  \BibitemOpen
  \bibfield  {author} {\bibinfo {author} {\bibfnamefont {Y.~N.}\ \bibnamefont
  {Obukhov}}\ and\ \bibinfo {author} {\bibfnamefont {D.}~\bibnamefont
  {Puetzfeld}},\ }\href {\doibase 10.1103/PhysRevD.90.024004} {\bibfield
  {journal} {\bibinfo  {journal} {Phys. Rev. D}\ }\textbf {\bibinfo {volume}
  {90}},\ \bibinfo {pages} {024004} (\bibinfo {year} {2014})},\ \Eprint
  {http://arxiv.org/abs/1405.4003} {arXiv:1405.4003 [gr-qc]} \BibitemShut
  {NoStop}%
\bibitem [{\citenamefont {Obukhov}\ \emph {et~al.}(2015)\citenamefont
  {Obukhov}, \citenamefont {Portales-Oliva}, \citenamefont {Puetzfeld},\ and\
  \citenamefont {Rubilar}}]{Obukhovetal2015}%
  \BibitemOpen
  \bibfield  {author} {\bibinfo {author} {\bibfnamefont {Y.~N.}\ \bibnamefont
  {Obukhov}}, \bibinfo {author} {\bibfnamefont {F.}~\bibnamefont
  {Portales-Oliva}}, \bibinfo {author} {\bibfnamefont {D.}~\bibnamefont
  {Puetzfeld}}, \ and\ \bibinfo {author} {\bibfnamefont {G.~F.}\ \bibnamefont
  {Rubilar}},\ }\href {\doibase 10.1103/PhysRevD.92.104010} {\bibfield
  {journal} {\bibinfo  {journal} {Phys. Rev. D}\ }\textbf {\bibinfo {volume}
  {92}},\ \bibinfo {pages} {104010} (\bibinfo {year} {2015})},\ \Eprint
  {http://arxiv.org/abs/1507.02191} {arXiv:1507.02191 [gr-qc]} \BibitemShut
  {NoStop}%
\bibitem [{\citenamefont {Schmidt}\ and\ \citenamefont
  {Bi\ifmmode~\check{c}\else \v{c}\fi{}\'ak}(2018)}]{SchmidtBicak2018}%
  \BibitemOpen
  \bibfield  {author} {\bibinfo {author} {\bibfnamefont {J.}~\bibnamefont
  {Schmidt}}\ and\ \bibinfo {author} {\bibfnamefont {J.}~\bibnamefont
  {Bi\ifmmode~\check{c}\else \v{c}\fi{}\'ak}},\ }\href {\doibase
  10.1063/1.5003190} {\bibfield  {journal} {\bibinfo  {journal} {J. Math. Phys.
  (N.Y.)}\ }\textbf {\bibinfo {volume} {59}},\ \bibinfo {pages} {042501}
  (\bibinfo {year} {2018})},\ \Eprint {http://arxiv.org/abs/1804.02298}
  {arXiv:1804.02298 [gr-qc]} \BibitemShut {NoStop}%
\bibitem [{\citenamefont {Iyer}\ and\ \citenamefont
  {Wald}(1994)}]{Iyer:1994ys}%
  \BibitemOpen
  \bibfield  {author} {\bibinfo {author} {\bibfnamefont {V.}~\bibnamefont
  {Iyer}}\ and\ \bibinfo {author} {\bibfnamefont {R.~M.}\ \bibnamefont
  {Wald}},\ }\href {\doibase 10.1103/PhysRevD.50.846} {\bibfield  {journal}
  {\bibinfo  {journal} {Phys. Rev. D}\ }\textbf {\bibinfo {volume} {50}},\
  \bibinfo {pages} {846} (\bibinfo {year} {1994})},\ \Eprint
  {http://arxiv.org/abs/gr-qc/9403028} {arXiv:gr-qc/9403028} \BibitemShut
  {NoStop}%
\bibitem [{\citenamefont {Jacobson}\ and\ \citenamefont
  {Mohd}(2015)}]{Jacobson:2015uqa}%
  \BibitemOpen
  \bibfield  {author} {\bibinfo {author} {\bibfnamefont {T.}~\bibnamefont
  {Jacobson}}\ and\ \bibinfo {author} {\bibfnamefont {A.}~\bibnamefont
  {Mohd}},\ }\href {\doibase 10.1103/PhysRevD.92.124010} {\bibfield  {journal}
  {\bibinfo  {journal} {Phys. Rev. D}\ }\textbf {\bibinfo {volume} {92}},\
  \bibinfo {pages} {124010} (\bibinfo {year} {2015})},\ \Eprint
  {http://arxiv.org/abs/1507.01054} {arXiv:1507.01054 [gr-qc]} \BibitemShut
  {NoStop}%
\bibitem [{\citenamefont {Prabhu}(2017)}]{Prabhu:2015vua}%
  \BibitemOpen
  \bibfield  {author} {\bibinfo {author} {\bibfnamefont {K.}~\bibnamefont
  {Prabhu}},\ }\href {\doibase 10.1088/1361-6382/aa536b} {\bibfield  {journal}
  {\bibinfo  {journal} {Class. Quant. Grav.}\ }\textbf {\bibinfo {volume}
  {34}},\ \bibinfo {pages} {035011} (\bibinfo {year} {2017})},\ \Eprint
  {http://arxiv.org/abs/1511.00388} {arXiv:1511.00388 [gr-qc]} \BibitemShut
  {NoStop}%
\bibitem [{\citenamefont {Chakraborty}\ and\ \citenamefont
  {Dey}(2018)}]{Chakraborty:2018qew}%
  \BibitemOpen
  \bibfield  {author} {\bibinfo {author} {\bibfnamefont {S.}~\bibnamefont
  {Chakraborty}}\ and\ \bibinfo {author} {\bibfnamefont {R.}~\bibnamefont
  {Dey}},\ }\href {\doibase 10.1016/j.physletb.2018.10.027} {\bibfield
  {journal} {\bibinfo  {journal} {Phys. Lett. B}\ }\textbf {\bibinfo {volume}
  {786}},\ \bibinfo {pages} {432} (\bibinfo {year} {2018})},\ \Eprint
  {http://arxiv.org/abs/1806.05840} {arXiv:1806.05840 [gr-qc]} \BibitemShut
  {NoStop}%
\bibitem [{\citenamefont {Aneesh}\ \emph {et~al.}(2020)\citenamefont {Aneesh},
  \citenamefont {Chakraborty}, \citenamefont {Hoque},\ and\ \citenamefont
  {Virmani}}]{Aneesh:2020fcr}%
  \BibitemOpen
  \bibfield  {author} {\bibinfo {author} {\bibfnamefont {P.}~\bibnamefont
  {Aneesh}}, \bibinfo {author} {\bibfnamefont {S.}~\bibnamefont {Chakraborty}},
  \bibinfo {author} {\bibfnamefont {S.~J.}\ \bibnamefont {Hoque}}, \ and\
  \bibinfo {author} {\bibfnamefont {A.}~\bibnamefont {Virmani}},\ }\href
  {\doibase 10.1088/1361-6382/aba5ab} {\bibfield  {journal} {\bibinfo
  {journal} {Class. Quant. Grav.}\ }\textbf {\bibinfo {volume} {37}},\ \bibinfo
  {pages} {205014} (\bibinfo {year} {2020})},\ \Eprint
  {http://arxiv.org/abs/2004.10215} {arXiv:2004.10215 [hep-th]} \BibitemShut
  {NoStop}%
\bibitem [{\citenamefont {Padmanabhan}(2014)}]{Padmanabhan:2013nxa}%
  \BibitemOpen
  \bibfield  {author} {\bibinfo {author} {\bibfnamefont {T.}~\bibnamefont
  {Padmanabhan}},\ }\href {\doibase 10.1007/s10714-014-1673-7} {\bibfield
  {journal} {\bibinfo  {journal} {Gen. Rel. Grav.}\ }\textbf {\bibinfo {volume}
  {46}},\ \bibinfo {pages} {1673} (\bibinfo {year} {2014})},\ \Eprint
  {http://arxiv.org/abs/1312.3253} {arXiv:1312.3253 [gr-qc]} \BibitemShut
  {NoStop}%
\bibitem [{\citenamefont {Chakraborty}\ and\ \citenamefont
  {Padmanabhan}(2015)}]{Chakraborty:2015hna}%
  \BibitemOpen
  \bibfield  {author} {\bibinfo {author} {\bibfnamefont {S.}~\bibnamefont
  {Chakraborty}}\ and\ \bibinfo {author} {\bibfnamefont {T.}~\bibnamefont
  {Padmanabhan}},\ }\href {\doibase 10.1103/PhysRevD.92.104011} {\bibfield
  {journal} {\bibinfo  {journal} {Phys. Rev. D}\ }\textbf {\bibinfo {volume}
  {92}},\ \bibinfo {pages} {104011} (\bibinfo {year} {2015})},\ \Eprint
  {http://arxiv.org/abs/1508.04060} {arXiv:1508.04060 [gr-qc]} \BibitemShut
  {NoStop}%
\bibitem [{\citenamefont {Chakraborty}(2015)}]{Chakraborty:2015wma}%
  \BibitemOpen
  \bibfield  {author} {\bibinfo {author} {\bibfnamefont {S.}~\bibnamefont
  {Chakraborty}},\ }\href {\doibase 10.1007/JHEP08(2015)029} {\bibfield
  {journal} {\bibinfo  {journal} {JHEP}\ }\textbf {\bibinfo {volume} {08}},\
  \bibinfo {pages} {029} (\bibinfo {year} {2015})},\ \Eprint
  {http://arxiv.org/abs/1505.07272} {arXiv:1505.07272 [gr-qc]} \BibitemShut
  {NoStop}%
\bibitem [{Mat()}]{MathematicaRef}%
  \BibitemOpen
  \href@noop {} {}\bibinfo {note}
  {\url{https://github.com/justincfeng/mathematica-files/blob/a7f627be270e2de69ea9e82edbbe6a1e3e640d8b/2020/AKEPert.nb}}\BibitemShut
  {NoStop}%
\bibitem [{\citenamefont {Feng}\ and\ \citenamefont
  {Matzner}(2018)}]{FengMatznerWeiss2018}%
  \BibitemOpen
  \bibfield  {author} {\bibinfo {author} {\bibfnamefont {J.~C.}\ \bibnamefont
  {Feng}}\ and\ \bibinfo {author} {\bibfnamefont {R.~A.}\ \bibnamefont
  {Matzner}},\ }\href {\doibase 10.1007/s10714-018-2420-2} {\bibfield
  {journal} {\bibinfo  {journal} {Gen. Relativ. Gravit.}\ }\textbf {\bibinfo
  {volume} {50}},\ \bibinfo {pages} {99} (\bibinfo {year} {2018})},\ \Eprint
  {http://arxiv.org/abs/1708.04489} {arXiv:1708.04489 [gr-qc]} \BibitemShut
  {NoStop}%
\bibitem [{\citenamefont {Sudarshan}\ and\ \citenamefont
  {Mukunda}(1983)}]{SudarshanCM}%
  \BibitemOpen
  \bibfield  {author} {\bibinfo {author} {\bibfnamefont {E.}~\bibnamefont
  {Sudarshan}}\ and\ \bibinfo {author} {\bibfnamefont {N.}~\bibnamefont
  {Mukunda}},\ }\href@noop {} {\emph {\bibinfo {title} {Classical Dynamics: A
  Modern Perspective}}}\ (\bibinfo  {publisher} {Krieger, Melbourne, FL},\
  \bibinfo {year} {1983})\BibitemShut {NoStop}%
\bibitem [{\citenamefont {Matzner}\ and\ \citenamefont
  {Shepley}(1991)}]{MatznerShepleyCM}%
  \BibitemOpen
  \bibfield  {author} {\bibinfo {author} {\bibfnamefont {R.~A.}\ \bibnamefont
  {Matzner}}\ and\ \bibinfo {author} {\bibfnamefont {L.~C.}\ \bibnamefont
  {Shepley}},\ }\href@noop {} {\emph {\bibinfo {title} {Classical Mechanics}}}\
  (\bibinfo  {publisher} {Prentice-Hall, Englewood Cliffs, NJ},\ \bibinfo
  {year} {1991})\BibitemShut {NoStop}%
\bibitem [{\citenamefont {Weiss}(1936)}]{Weiss1936}%
  \BibitemOpen
  \bibfield  {author} {\bibinfo {author} {\bibfnamefont {P.}~\bibnamefont
  {Weiss}},\ }\href {\doibase 10.1098/rspa.1936.0143} {\bibfield  {journal}
  {\bibinfo  {journal} {Proc. R. Soc. Lond. A}\ }\textbf {\bibinfo {volume}
  {156}},\ \bibinfo {pages} {192} (\bibinfo {year} {1936})}\BibitemShut
  {NoStop}%
\bibitem [{\citenamefont {Gundlach}\ and\ \citenamefont
  {Mart{\'{\i}}n-Garc{\'{\i}}a}(2006)}]{Gun06}%
  \BibitemOpen
  \bibfield  {author} {\bibinfo {author} {\bibfnamefont {C.}~\bibnamefont
  {Gundlach}}\ and\ \bibinfo {author} {\bibfnamefont {J.~M.}\ \bibnamefont
  {Mart{\'{\i}}n-Garc{\'{\i}}a}},\ }\href {\doibase
  10.1088/0264-9381/23/16/s06} {\bibfield  {journal} {\bibinfo  {journal}
  {Classical Quantum Gravity}\ }\textbf {\bibinfo {volume} {23}},\ \bibinfo
  {pages} {S387} (\bibinfo {year} {2006})},\ \Eprint
  {http://arxiv.org/abs/gr-qc/0506037} {arXiv:gr-qc/0506037} \BibitemShut
  {NoStop}%
\bibitem [{\citenamefont {Hilditch}(2013)}]{Hil13}%
  \BibitemOpen
  \bibfield  {author} {\bibinfo {author} {\bibfnamefont {D.}~\bibnamefont
  {Hilditch}},\ }\href {\doibase 10.1142/S0217751X13400150} {\bibfield
  {journal} {\bibinfo  {journal} {Int. J. Mod. Phys.}\ }\textbf {\bibinfo
  {volume} {A28}},\ \bibinfo {pages} {1340015} (\bibinfo {year} {2013})},\
  \Eprint {http://arxiv.org/abs/1309.2012} {arXiv:1309.2012 [gr-qc]}
  \BibitemShut {NoStop}%
\bibitem [{\citenamefont {Lynden-Bell}\ and\ \citenamefont
  {Bi\ifmmode~\check{c}\else \v{c}\fi{}\'ak}(2017)}]{Lynden-BellBicak2017}%
  \BibitemOpen
  \bibfield  {author} {\bibinfo {author} {\bibfnamefont {D.}~\bibnamefont
  {Lynden-Bell}}\ and\ \bibinfo {author} {\bibfnamefont {J.}~\bibnamefont
  {Bi\ifmmode~\check{c}\else \v{c}\fi{}\'ak}},\ }\href {\doibase
  10.1103/PhysRevD.96.104053} {\bibfield  {journal} {\bibinfo  {journal} {Phys.
  Rev. D}\ }\textbf {\bibinfo {volume} {96}},\ \bibinfo {pages} {104053}
  (\bibinfo {year} {2017})},\ \Eprint {http://arxiv.org/abs/1712.06980}
  {arXiv:1712.06980 [gr-qc]} \BibitemShut {NoStop}%
\bibitem [{\citenamefont {Chakraborty}\ and\ \citenamefont
  {Padmanabhan}(2020)}]{Chakraborty:2019doh}%
  \BibitemOpen
  \bibfield  {author} {\bibinfo {author} {\bibfnamefont {S.}~\bibnamefont
  {Chakraborty}}\ and\ \bibinfo {author} {\bibfnamefont {T.}~\bibnamefont
  {Padmanabhan}},\ }\href {\doibase 10.1103/PhysRevD.101.064023} {\bibfield
  {journal} {\bibinfo  {journal} {Phys. Rev. D}\ }\textbf {\bibinfo {volume}
  {101}},\ \bibinfo {pages} {064023} (\bibinfo {year} {2020})},\ \Eprint
  {http://arxiv.org/abs/1909.00096} {arXiv:1909.00096 [gr-qc]} \BibitemShut
  {NoStop}%
\bibitem [{\citenamefont {Mart{\'i}n-Garc{\'i}a}(2020)}]{xActbib}%
  \BibitemOpen
  \bibfield  {author} {\bibinfo {author} {\bibfnamefont {J.~M.}\ \bibnamefont
  {Mart{\'i}n-Garc{\'i}a}},\ }\href@noop {} {\enquote {\bibinfo {title}
  {x{A}ct: tensor computer algebra.}}\ } (\bibinfo {year} {2020}),\ \bibinfo
  {note} {\url{http://www.xact.es/}}\BibitemShut {NoStop}%
\end{thebibliography}%

%=======================================================================

\end{document}